\documentclass[aps,pra,twocolumn,floatfix,superscriptaddress]{revtex4-1}

\usepackage{amsmath}

\usepackage{subfigure}
\usepackage{graphicx}
\usepackage[T1]{fontenc}
\usepackage{ulem}
\usepackage{xcolor}

\begin{document}

\title{Mass-imbalance induced structures of binary atomic mixtures in box potentials}
\date{\today}

\author{Bishal Parajuli}
\affiliation{Department of Physics, University of California, Merced, CA 95343, USA.}

\author{Daniel P{\k e}cak}
\affiliation{\mbox{Faculty of Physics, Warsaw University of Technology, Ulica Koszykowa 75, 00-662 Warsaw, Poland}}

\author{Chih-Chun Chien}
\affiliation{Department of Physics, University of California, Merced, CA 95343, USA.}
\email{cchien5@ucmerced.edu}

\begin{abstract}
We consider the ground states of binary atomic boson-boson and fermion-fermion mixtures confined in one-dimensional box potentials by simulating the systems using few-body models with delta-function interactions and many-body models with density-density interactions. For boson-boson mixtures, both models show signatures of phase separation in the strong repulsion regime and sandwiched structures emerge in the presence of mass imbalance. The structural difference between equal-mass and mass-imbalanced systems is due to the minimization of the interaction energy and the kinetic energies from the density distortion at the hard walls and at the phase-separation interface. The mass imbalance adjusts the kinetic energies and causes the lighter species to avoid the hard walls. For fermion-fermion mixtures, few-body simulations show a mass-imbalance induced structural changes in the strong repulsion regime, while many-body simulations show two-chunk phase separation due to the strong bulk kinetic energy. For equal-mass mixtures with strong inter-species repulsion, the few-body and many-body models predict different structures because the mean-field treatment in the many-body model approximates the contact interaction and smooths out the wavefunctions.
\end{abstract}

\maketitle

\section{Introduction}
Ultracold atoms trapped in optical potentials have been an important system for studying many-body quantum physics such as Bose-Einstein condensation (BEC) and superfluidity~\cite{pitaevskii2003bose,pethick2008bose,ueda2010fundamentals,stoof2008ultracold}.  However, conventional harmonic traps cause inhomogeneous density profiles and complicate comparisons between theories and experiments. For instance, BEC may emerge at the trap center while the normal phase still occupies the edge. To construct theories for harmonically trapped gases, one may use the local density approximation~\cite{pethick2008bose,ueda2010fundamentals} to map out the profiles of physical quantities before comparing the results to experiments.

Realizations of box potentials for ultracold atoms~\cite{Es10,Gaunt2013,PhysRevLett.112.040403,PhysRevLett.118.123401,Horikoshi17,PhysRevLett.120.060402} have brought possibilities of measuring homogeneous bulk properties directly. The introduction of hard-wall potentials from the laser sheets for generating the box potential imposes open boundary condition. On the other hand, one may approximate homogeneous systems with periodic boundary condition by trapping ultracold atoms in ring-shaped potentials~\cite{Wright11,Eckel14,PhysRevA.83.043408,PhysRevA.74.023617,Chen19}. One important difference between open and periodic boundary conditions is the kinetic energy caused by the vanishing of the wavefunction at the hard walls.

Mixtures of ultracold atoms in distinct internal states or from different species exhibit interesting structures and thermodynamic properties~\cite{pethick2008bose,ueda2010fundamentals}.
When a binary atomic mixture has strong inter-species repulsion, the two components tend to separate from each other and form phase separation as demonstrated in Ref.~\cite{PhysRevLett.81.1539}. In a phase-separation structure, the interface between the two species also creates additional kinetic energy because the density profiles changes drastically, which may be viewed as the surface tension of the interface~\cite{PhysRevLett.97.070402,trappe2016ground}.

Here we focus on binary mixtures and label the two species as 1 and 2. When the two species phase separate, a competition between the energy increase due to the distortion of wavefunctions near the hard-walls and at the phase-separation interface can lead to different structures of the mixture. Since the kinetic energy is inversely proportional to the mass, we will compare the structures with and without mass imbalance in the strong repulsion regime. For example, if the distortion of species 2 at the hard wall leads to a substantial kinetic energy increase compared to the interfacial energy between species 1 and 2, the system will form a sandwich structure with species 2 away from both hard-walls. We will show that the lighter species tend to avoid the hard walls. In contrast, equal-mass mixtures tend to remain miscible or separate into two chunks depending on how the interactions are modeled.

There have been theoretical studies on confinement effects~\cite{Garcia-MarchPRA2014,Volosniev14,Bellotti2017,dehkharghani_quantum_2015}, including hard-wall potentials~\cite{HaoPRA79_033607}, and mass effects~\cite{Iskin07,PhysRevLett.110.165302,Kang2018,takemori2012low,conduit2011itinerant,Han19,NJP013030} on the structures of atomic mixtures. In a recent review~\cite{Sowinski19}, a broad class of few-body systems in harmonic traps has been covered, including the mass-imbalanced systems discussed in Refs.~\cite{DehkharghaniPRA92_031601,MehtaPRA89_052706,MarchJPhysB49_075303,DehkharghaniJPhysB49_085301,VolosnievFewBodySyst,HarshmanPRX7_041001}. However, mass-imbalance induced structural changes due to hard-wall potentials have not been systematically investigated.  Studies on binary fermion mixtures have shown that repulsive interactions are not the only factor determining phase separation. Other factors such as population imbalance~\cite{Zwierlein492}, mass imbalance~\cite{PhysRevLett.110.165302,Kang2018,conduit2011itinerant,takemori2012low}, and additional p-wave interaction~\cite{Kang2018} also affect the structures. It has been proposed that phase separation of fermion mixtures in the thermodynamic limit driven by a large mass imbalance is possible in all dimensions even in the weakly interacting regime~\cite{PhysRevLett.110.165302}. Moreover, there are recent experiments on various atomic mixtures~\cite{cetina2016ultrafast,Grimm2018DyK,Grimm2018KLi,serwane2011deterministic}. We mention that the physics of atomic mixtures of fermions with strong repulsion is complex due to possible Stoner instability~\cite{Jo09} and formation of bound states~\cite{Pekker11}, but here we consider a simplified scenario where the fermionic mixtures exhibit phase separation in the strong-repulsion regime.

In ultra-cold atoms, one major difference between bosons and fermions is the lack of intra-species interactions between identical fermions. This is because Pauli exclusion principle suppresses the dominant, two-body s-wave interaction between identical fermions~\cite{pethick2008bose}. Another major difference between bosonic and fermionic mixtures comes from the Fermi pressure, which is related to the bulk kinetic energy and depends on the fermion density nonlinearly~\cite{Fetter_book}. The phase-separation structure needs to settle the competition between the interactions, kinetic energy due to the confinement, and the Fermi pressure in the case of fermion mixtures. Previous studies of fermionic mixtures in harmonically trapped gases \cite{trappe2016ground} have shown good agreement with experiments \cite{valtolina2017exploring}. Note that the single-particle energies in a harmonic trap grow linearly with the quantum number, but in a box trap they scale with the square of the quantum number~\cite{Schiff_book}. Thus, the Fermi pressure is more significant in a box trap potential, making it harder to reach spatial separation.

To investigate mass effects on the structures of atomic mixtures in box potentials, we implement both the few-body exact-diagonalization method~\cite{NJP013030,Pecak2016Transition,Sowinski19} as well as the many-body Gross-Pitaevskii and Hartree-Fock theories~\cite{Fetter_book,pitaevskii2003bose,Minguzzi04}. We will present structures of boson-boson and fermion-fermion mixtures with strong inter-species repulsion confined in a 1D box potential with two hard wall in the $x$ direction. Importantly, the wavefunctions have to vanish at the hard walls. The systems are assumed to be homogeneous in the $y$ and $z$ directions. 
Since thermal excitations have been shown to be negligible in the box-potential experiments at low temperatures~\cite{Es10,Gaunt2013,PhysRevLett.112.040403,PhysRevLett.118.123401,Horikoshi17}, we will focus on the ground-state properties of the systems. In the strong repulsion limit, the ground state may be quasi-degenerate~\cite{PyzhNJP20_015006} and additional analyses will be needed.

The contact interactions between atoms are modeled differently in the literature of few-body and many-body theories: In the few-body approach, the interaction corresponds to delta-function terms following the Lieb-Liniger model~\cite{LLmodel1,LLmodel2,Volosniev14,Pecak2016Transition}.
On the other hand, the delta pseudopotential may be approximated at mean-field level in many-body theories, leading to the density-density terms in the many-body treatments~\cite{Fetter_book,pitaevskii2003bose,pethick2008bose}.
Bosons with attractive interactions at low temperatures collapse as the interaction energy overcomes the zero point motion induced by the confinement~\cite{PhysRevLett.82.876,donley_dynamics_2001} while two-component fermions with attractive interactions may form Cooper pairs~\cite{pethick2008bose,ueda2010fundamentals}. Here we only consider bosons and fermions with repulsive interactions. In addition to the mean-field and exact-diagonalization methods used here, the multi-configurational methods~\cite{ZollnerPRA74_063611,Gwak18,ErnstPRA84_023623} have been  applied to cold-atom systems recently.

The structures from the few-body and many-body calculations agree in the case of boson-boson mixtures with large mass-imbalance and strong repulsion, where similar sandwich structures are predicted. Otherwise, the structures are different: (1) For equal-mass boson-boson or fermion-fermion mixtures in the strong-repulsion regime, the few-body calculation predicts a miscible phase with strong internal correlations while the many-body calculation predicts phase separation into two chunks.
(2) For fermion-fermion mixtures with large mass imbalance and strong inter-species repulsion, the many-body model predicts the two species separate into two chunks while the few-body model predicts a three-chunk sandwich structure.

By comparing the structures with and without mass imbalance, we found mass-imbalance induced structural changes when comparing the equal-mass and highly mass-imbalanced cases: (1) The few-body calculations of bosonic mixtures predict a change from a miscible phase to a sandwich structure. (2) The many-body calculations of bosonic mixtures predict a change from two-chunk phase separation to a sandwich structure. (3) The few-body calculations of fermionic mixtures predict a change from a mixture to a sandwich structure. On the other hand, the many-body calculation predicts a two-chunk phase separation structure of fermionic mixtures that is insensitive to mass imbalance because of the strong bulk kinetic energy in the many-body system.

The rest of the paper is organized as follows. Section~\ref{sec:Bmixture} presents the structures of boson-boson mixtures in a 1D box potential from few-body and many-body theories. Mass-imbalance induced structural changes are found in both cases. Section~\ref{sec:Fmixture} presents the results of fermion-fermion mixtures in a 1D box potential from few-body and many-body theories. Only the few-body results predict a mass-imbalance induced structural change. Section~\ref{sec:exp} summarizes some implications for experiments. Section~\ref{sec:Conclusion} concludes our work.

\section{Boson-boson mixtures in a box potential}\label{sec:Bmixture}
\subsection{Few-body theory and result}
Here we present our investigation of a few-body boson-boson mixture of equal number of particles $N_1=N_2=N$ in a 1D box potential with hard walls at $x=0, L$.
The spatial separation of a fermion-fermion mixture in a box potential was recently studied by few-body calculations~\cite{Pecak2016Transition}. The mechanism of the phase separation works in principle for different statistics, including the bosonic mixtures studied here.

The contact interaction in the few-body calculation is modeled by a $\delta$-function pseudo-potential similar to the Lieb-Liniger model~\cite{LLmodel1,LLmodel2}. 
We assume the intra-species interactions are negligible compared to the inter-species interactions. Although including the intra-species interactions in our study only leads to quantitative differences, richer physics may happen when the interactions are in the ultra-strong regime~\cite{ZollnerPRA78_013629,HaoEPJD51_261,PyzhNJP20_015006}.
The boson-boson Hamiltonian of the system studied is given by
\begin{multline}\label{eq:ham}
 {\cal \hat{H}} = \sum_{\alpha=1}^{2} \int_0^L \!\mathrm{d}x 
 \hat{\Psi}_\alpha^\dag(x) H_\alpha \hat{\Psi}_\alpha(x) \\
 + g_{12} \int_0^L \!\mathrm{d}x  \hat{\Psi}_1^\dag(x)\hat{\Psi}_2^\dag(x) \hat{\Psi}_2(x) \hat{\Psi}_1(x),
\end{multline}
where the single-particle Hamiltonians contain only the kinetic part:
\begin{equation}\label{eq:spHam}
 H_\alpha = -\frac{\hbar^2}{2m_\alpha} \frac{\partial^2}{\partial x^2}.
\end{equation}
The integrations are over the whole system of length $L$, which sets the length scale.
The effective one-dimensional inter-species interaction $g_{12}$  can be obtained from the three-dimensional scattering length by integrating out two dimensions~\cite{Olshanii1998}.
The field operator $\hat{\Psi}_\alpha(x)$ annihilates a boson of species $\alpha$ and mass $m_\alpha$ at position $x$ and obeys the standard bosonic commutation relation: $\left[\hat{\Psi}_\alpha^\dag(x),\hat{\Psi}_\beta(x')\right]=\delta(x-x')\delta_{\alpha\beta}$.
The operators of different species commute.
For convenience we will denote the mass ratio by $\mu=m_2/m_1$.

\subsubsection{Method}
To obtain the ground state and to study its properties, we expand the field operators in the single-particle basis $\phi_{\alpha i}(x)$ of a box potential for a given component $\alpha$:
\begin{equation}
 \hat{\Psi}_\alpha(x) = \sum_i \phi_{\alpha i}(x) \hat{b}_{\alpha i}.
\end{equation}
Then the Hamiltonian reads:
\begin{equation}\label{eq:sqHam}
 {\cal \hat{H}} = \sum_{\alpha=1}^{2} \sum_i E_{\alpha i}
 \hat{b}_{\alpha i}^\dag \hat{b}_{\alpha i}
 + g_{12} \sum_{uvkl} U_{uvkl} \hat{b}^\dag_{1l}\hat{b}^\dag_{2k}\hat{b}_{2v}\hat{b}_{1u},
\end{equation}
where $E_{\alpha i}$ is the $i$-th single-particle energy of species $\alpha$ and the interaction is proportional with $g_{12}$ to the term that is defined as 
\begin{equation}
 U_{uvkl} = \int_0^L \!\mathrm{d}x \phi^*_{1l}(x)\phi^*_{2k}(x)\phi_{2v}(x)\phi_{1u}(x).
\end{equation}
Here, due to the bosonic enhancement, $g_{12}=1$ corresponds to the regime where the interaction is comparable to the kinetic energy.

The sum in Eq.~\eqref{eq:sqHam} contains infinite number of terms. However, to tackle the problem numerically one should limit the size of the Hilbert space, for example by setting a cutoff on the number of single-particle states.
The size of the Hilbert space grows rapidly, and it contains the states of different orders of single-particle energies.
To reduce the size of the Hilbert space, we include all possible states up to a certain energy cutoff $E_{MAX}$~\cite{plodzien2018numerically}.
We show the size of the Hilbert space for binary boson-boson (fermion-fermion) mixtures in Table~\ref{tabbb} (Table~\ref{tabff}). The complexity grows exponentially with the number of particles and single-particle eigenstates, setting a limitation of the few-body calculations.
The convergence of the few-body calculations can be checked in many complementary ways such as overlap between wave functions or two-body correlations \cite{Pecak2017com,Pecak2019Corelation}. In our simulations, we have checked that the single-particle densities do not change significantly as the many-body cutoff in energy is varied (see Appendix \ref{appendix}). The satisfaction of this condition is highly dependent on the number of particles, statistics, and interaction strength, and it should be adjusted individually for every set of parameters. 

\begin{table}
\caption{Size of the Hilbert space for $N$ bosons of each type and $k$ single-particle orbitals.}
\begin{tabular}{c||c|c|c|c}
    $k$ | $N$ & 3 & 4 & 5 & 6 \\ \hline \hline 
    10        &  48,400 &   511,225 &  4,008,004 &  25,050,025 \\ \hline
    11        &  81,796 & 1,002,001 &  9,018,009 &  64,128,064 \\ \hline
    12        & 132,496 & 1,863,225 & 19,079,424 & 153,165,376 \\ \hline
    13        & 207,025 & 3,312,400 & 38,291,344 & 344,622,096 \\ \hline
    14        & 313,600 & 5,664,400 & 73,410,624 & 736,145,424 \\ \hline
\end{tabular}
\label{tabbb}
\end{table}

\begin{table}
\caption{Size of the Hilbert space for $N$ fermions of each type and $k$ single-particle orbitals.}
\begin{tabular}{c||c|c|c|c}
    $k$ | $N$ & 3 & 4 & 5 & 6 \\ \hline \hline 
    10        &  14,400 &    44,100 &    63,504 &    44,100  \\ \hline
    11        &  27,225 &   108,900 &   213,444 &   213,444 \\ \hline
    12        &  48,400 &   245,025 &   627,264 &   853,776 \\ \hline
    13        &  81,796 &   511,225 & 1,656,369 & 2,944,656 \\ \hline
    14        & 132,496 & 1,002,001 & 4,008,004 & 9,018,009 \\ \hline
\end{tabular}
\label{tabff}
\end{table}

Subsequently, we perform  exact diagonalization by using Arnoldi method for sparse matrices~\cite{ARPACK1998Sorensen}, which gives access to the ground state and its energy.
The wavefunction is then used to obtain the observables of interest.
The single-particle density $\rho_\alpha(x)$ for a given species $\alpha$ is defined for the ground state $|G_0\rangle$ of a system as
\begin{equation}
 \rho_\alpha(x) = \langle G_0|\hat{\Psi}_\alpha^\dag(x)\hat{\Psi}_\alpha(x)|G_0\rangle.
\end{equation}

\subsubsection{Equal-mass mixture}
\begin{figure}[t]
 \includegraphics[width=\columnwidth]{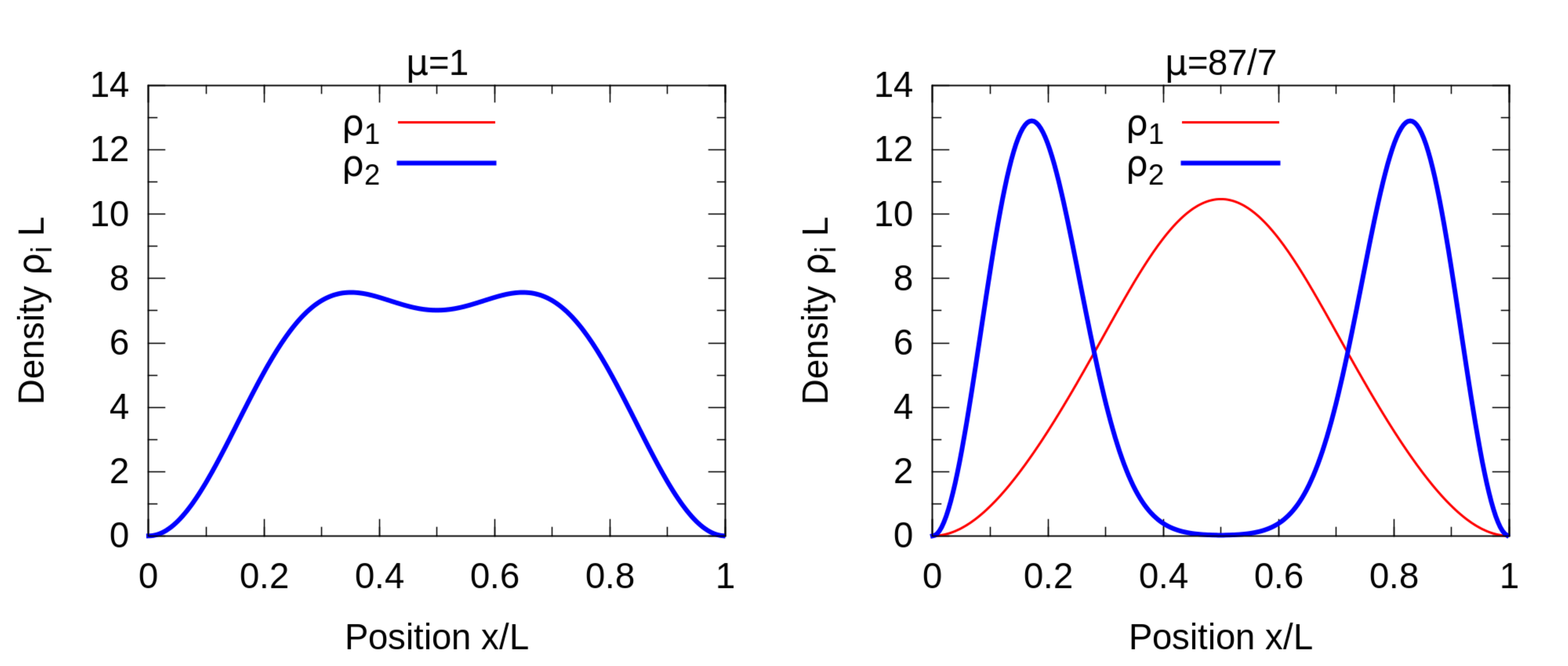}
 \caption{\label{fig:fewBosons}
 Single-particle density profiles of a system of $N=5$ bosons of each type. Here the interspecies repulsion strength is $g_{12}=1$ in the strongly interacting regime. The left and right panels show the equal mass case with $\mu=1$ and the $^{87}$Rb-$^{7}$Li mixture case ($\mu=m_2/m_1=87/7$), respectively.
 }
\end{figure}

The single-particle density profile for an equal-mass mixture of $N=5$ particles of each type is shown in the left panel of Fig.~\ref{fig:fewBosons}. Due to the symmetry between the two species in the Hamiltonian, the equal mass case does not distinguish between the species and the two density profiles coincide with each other.
From a practical point of view, the result shown in Fig.~\ref{fig:fewBosons} is the expectation value averaged over many measurements. In reality, the measurement of the system from shot to shot may show deviations. Moreover, our simplified model does not take into account complications from system-reservoir coupling or imbalance between the numbers of particles that may serve as a symmetry breaking mechanisms.

The perfect match of the single-particle densities of the two species does not reveal information about the correlations between the bosons.
However, there should be correlations between bosons of the same species due to the Bose distribution and between different species due to the inter-species repulsion. To analyze the correlations, we study the interspecies and intraspecies two-body correlations defined on the ground state $|G_0\rangle$ as ($\alpha,\beta=1,2$)
\begin{equation}\label{eq:Cab}
 C_{\alpha\beta}(x_1,x_2) = \langle G_0|\hat{\Psi}_\alpha^\dag(x_1)\hat{\Psi}_\beta^\dag(x_2)\hat{\Psi}_\beta(x_2)\hat{\Psi}_\alpha(x_1)|G_0\rangle.
\end{equation}
Note that due to the symmetry between species $1$ and $2$ in the equal-mass case, $C_{11}(x_1,x_2)=C_{22}(x_1,x_2)$.

\begin{figure}[t]
 \includegraphics[width=\columnwidth]{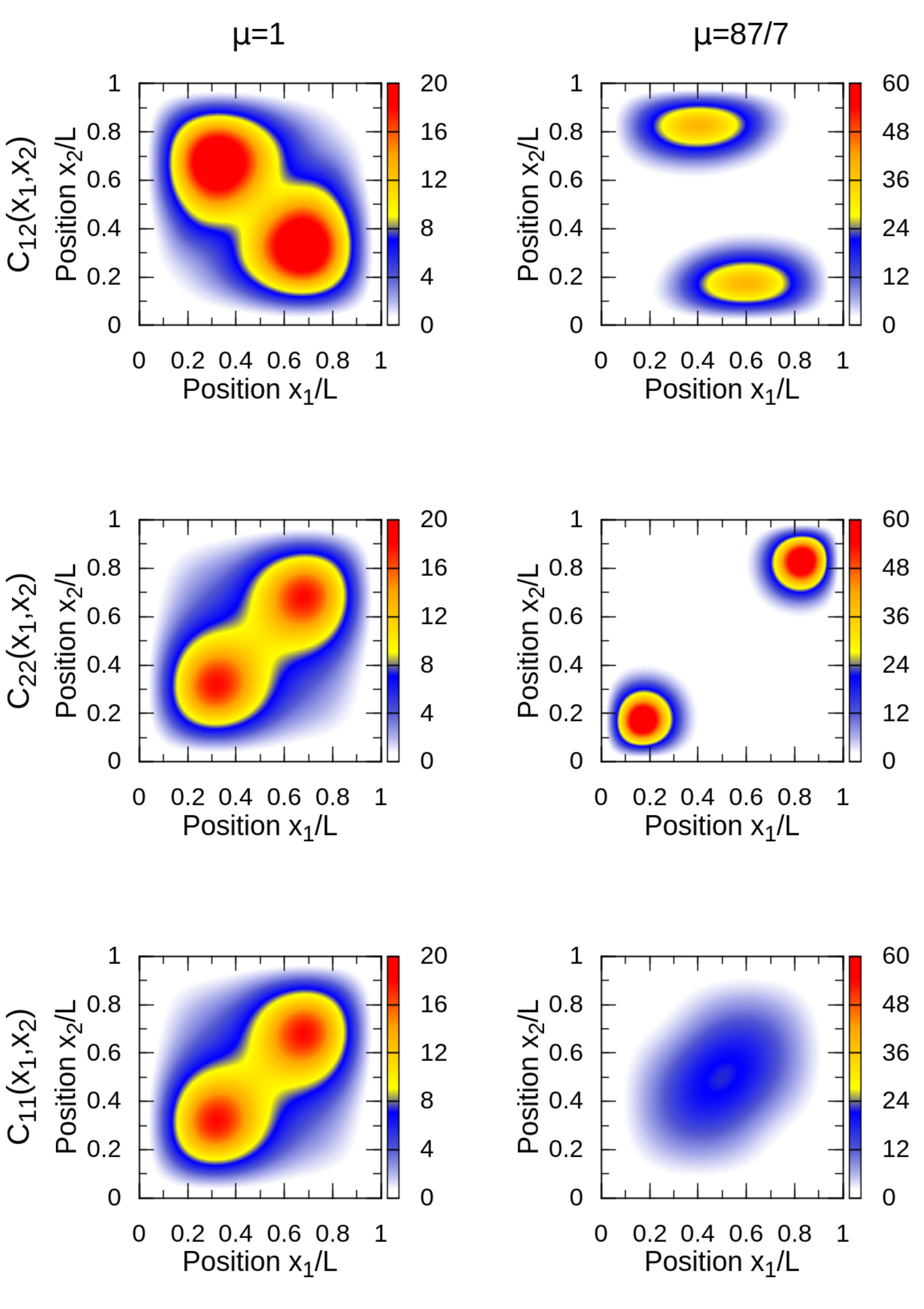}
 \caption{\label{fig:fewBosonsCorr}
 Two-particle density correlations $C_{\alpha\beta}$ defined in Eq.~\eqref{eq:Cab} for binary bosonic mixtures with $N=5$ particles of each type. Here $x_1,x_2$ label the positions on the $x$-axis of the box. The interspecies repulsion is $g_{12}=1$ in the strong repulsion regime. The left (right) column shows the equal-mass case with mass ratio $\mu=1$ ($^{87}$Rb-$^7$Li mixture with $\mu=87/7$).
 }
\end{figure}

As shown in the left column of Fig.~\ref{fig:fewBosonsCorr}, the two-body correlations reveal the ferromagnetic structure of the system, where the correlations congregate into two chunks.
The correlations imply that when a measurement of the densities is performed on the system, one species will tend to occupy the left part of the box while the other species will occupy the right. The average over many measurements, however, reveal the same averaged density profiles for both species. The ferromagnetic behavior is consistent with the many-body result that will be shown later.

\subsubsection{Mass-imbalanced mixture}
On the right panel of Fig.~\ref{fig:fewBosons}, the single-particle densities are presented for a $^{87}$Rb-$^{7}$Li mixture with mass ratio $\mu=87/7$.
In this mixture, the symmetry between the two species is broken by the mass imbalance. According to Eq.~\eqref{eq:spHam}, the single-particle energies depend explicitly on the mass, namely the excitation of the heavy particles cost less energy than the excitation of the light particles.
The correlations shown in the right column of Fig.~\ref{fig:fewBosonsCorr} corroborate the phase separation structure. 
We have checked, similar to Ref.~\cite{GirardeauPRA63_033601,MarchPRA88_063604}, how many natural orbitals are needed to properly describe the system. Explicitly, we diagonalized the single-particle density matrix for each component and compared the eigenvalues corresponding to the natural orbitals. We found the light component in the central region can be well described by just one natural orbital while the heavy component on the two sides needs two natural orbitals.

The mass imbalance introduces a mechanism that lowers significantly the inter-species interaction energy by phase separation. Although the local kinetic energy increases in each chunk, different species avoid each other and reduce the interaction energy. Moreover, the density of the light species vanishes next to the hard walls while the heavy species reside there. The sandwich structure further lowers the kinetic energies due to distortion of the wavefunctions.

We note that phase separation often occurs when an asymmetry is introduced. For example,  mass imbalance differentiates the two species and changes the single-particle energies in a box potential~\cite{Pecak2016Transition} or harmonic trap~\cite{NJP013030}. Moreover, different intra-species interactions $g_{11} \neq g_{22}$ may also be used~\cite{MarchPRA88_063604}.

\subsection{Many-body theory and result}
Here we use the Gross-Pitaevskii (GP) theory of bosonic condensate~\cite{pethick2008bose,pitaevskii2003bose} to study the structures of boson-boson mixtures in a 1D box potential. The GP theory has been mentioned in Ref.~\cite{PhysRevLett.81.1539} showing experimental data of harmonically trapped two-component bosons in the miscible phase and phase separation. Thus, we focus on the weak and intermediate interaction regimes where the GP theory works reasonably.

\subsubsection{Imaginary time formalism}
The Gross-Pitaevskii energy functional of a boson-boson mixture is~\cite{pethick2008bose}
\begin{align}
E=\int d^d x & \Big[\frac{\hbar^2}{2m_1}|\nabla\psi_1|^2+V_1(r)|\psi_1|^2 +\frac{\hbar^2}{2m_2}|\nabla\psi_2|^2
 \nonumber \\
&+V_2(r)|\psi_2|^2
+\frac{1}{2}g_{11}N_1^2|\psi_1|^4+\frac{1}{2}g_{22}N_2^2|\psi_2|^4
 \nonumber \\
&+g_{12}N_1N_2|\psi_1|^2|\psi_2|^2\Big].
\end{align}
Here we consider a mixture confined by two hard-walls at $x=0,L$ modeled by the potential $V_{\alpha}(r)$, and the system is uniform in the other directions. The condensate wavefunctions are normalized by $\int_0^L dx |\psi_\alpha|^2 =1$, where $\alpha=1,2$ denotes components. The mass $m_\alpha$ and number of particles $N_\alpha$ correspond to species $\alpha$.
The coupling constants $g_{11}$, $g_{22}$ and $g_{12}=g_{21}$ are related to the two-body $s$-wave scattering lengths $a_{11}$, $a_{22}$ and $a_{12}$ by
\begin{equation}
    g_{\alpha\beta}=\frac{2\pi \hbar^2 a_{\alpha\beta}}{m_{\alpha\beta}}.
\end{equation}   
Here  $m_{\alpha\beta}=\frac{m_\alpha m_\beta}{m_\alpha+m_\beta}$
is the reduced mass of pair of atoms and $\alpha,\beta=1,2$ denote the species. We focus on the repulsive case with $g_{\alpha\beta}>0$.  

The minimization of the energy functional, $\delta E/\delta \psi_{\alpha}^*=0$ for $\alpha=1,2$, leads to the time-independent Gross-Pitaevskii equation~\cite{pitaevskii2003bose}. To find the minimal-energy configuration starting from a given initial configuration, we implement the imaginary-time formalism~\cite{Fetter_book,pethick2008bose} by searching for the solution to the imaginary-time evolution equations $-\partial \psi_{\alpha}/\partial \tau=\delta E/\delta \psi_{\alpha}^*$ in the $\tau\rightarrow \infty$ limit with the normalization $\int|\psi_\alpha|^2 dx =1$ imposed at each imaginary-time increment. Here $\tau=it$ is the imaginary time. The mechanism behind the imaginary time evolution is that an arbitrary initial state can be decomposed by the many-body energy eigenstates by $\psi_{j}^{0}=\sum_{n}\psi_n \exp(-\beta E_n)$. Here $\beta=1/(k_B T)$. As $T\rightarrow 0$, the solution of the imaginary-time evolution in the $\tau\rightarrow\infty$ limit projects out the many-body ground state because the excited-state contributions decay away due to the normalization condition.

Therefore, we search for the solutions to the coupled imaginary-time evolution equations in the $x$-direction:
\begin{eqnarray}\label{eqn:gp1}
    -\hbar\frac{\partial \psi_1}{\partial \tau} &=& -\frac{\hbar^2}{2m_1}\partial_x^2\psi_1 + V_1\psi_1 + g_{11}N_1|\psi_1|^2\psi_1
     \nonumber \\
   && + g_{12}\sqrt{N_1 N_2}|\psi_2|^2\psi_1,  \nonumber \\
    -\hbar\frac{\partial \psi_2}{\partial \tau} &=& -\frac{\hbar^2}{2m_2}\partial_x^2\psi_2 + V_2\psi_2 + g_{22}N_2|\psi_2|^2\psi_2 
     \nonumber \\
    &&+ g_{12}\sqrt{N_1 N_2}|\psi_1|^2\psi_2.
\end{eqnarray}
The boundary conditions are $\psi_{\alpha}=0$ at $x=0,L$ for $\alpha=1, 2$.
We choose the units so that $\hbar=2m_1=1$.
The conservation of particle numbers imposes the following normalization condition of the density $\rho_\alpha=N_\alpha|\psi_\alpha|^2$. The total particle number of species $\alpha=1,2$ can then be obtained from
\begin{eqnarray}
N_{\alpha}=\int_0^L dx \rho_\alpha.
\end{eqnarray}

The GP equation provides an effective description of the macroscopic wavefunction of the BEC. It works well at low temperatures in the weakly interacting regime~\cite{pitaevskii2003bose,pethick2008bose}.
For a homogeneous, equal-mass boson-boson mixture with repulsive inter- and intra- species interactions in the thermodynamic limit, the  
stability condition has been summarized in Refs.~\cite{PhysRevLett.81.5718,PhysRevA.58.4836,pethick2008bose}. The system exhibits phase separation when 
$g_{11}g_{22}<g_{12}^2$.
Note that if any of the intra-species interaction vanishes, $g_{\alpha\alpha}=0$, the miscible phase is unstable against any finite inter-species interaction.
Here we will study how the mass imbalance affects the structure in the phase separation regime when the system is confined in a box potential. There have been other methods for obtaining density profiles of bosons or bosonic mixtures~\cite{KolomeiskyPRL85_1146,MarchPRA87_063633,TanatarPRA62_053601,GirardeauComment}, but here we use the GP equation because it reasonably describes the bosonic mixture experiment of Ref.~\cite{PhysRevLett.81.1539}.

We use the split step Crank-Nicolson method to solve the coupled imaginary-time evolution equations~\cite{MURUGANANDAM20091888,TAHA1984203}. The spatial and temporal increments are calibrated by checking against the exact solutions of nonlinear Schrodinger equations~\cite{0305-4470-35-47-309}.  In the following, we consider the structures of boson-boson mixtures with and without mass imbalance. For the mass-imbalanced case, we consider a mixture of $^7$Li and $^{87}$Rb, but our methods apply to other bosonic mixtures as well. We use the following parameters: 
$\Delta\tau= 0.0001\hbar/E_0$ for the imaginary time increment. $E_0=\hbar^2/(2m_1 L^2)$ is the energy unit. $N_1=N_2=100$ for the particle number of each species. By defining $g_0=4\pi\hbar^2L/m_1$, we express the coupling constants in terms of $g_{\alpha\beta}/g_0$. A $500$-point grid is used to discretize the space, and we have checked the results are insensitive to a further refinement of the grid.

\subsubsection{Results}
\begin{figure}
\subfigure[]{
\includegraphics[width=0.22\textwidth]{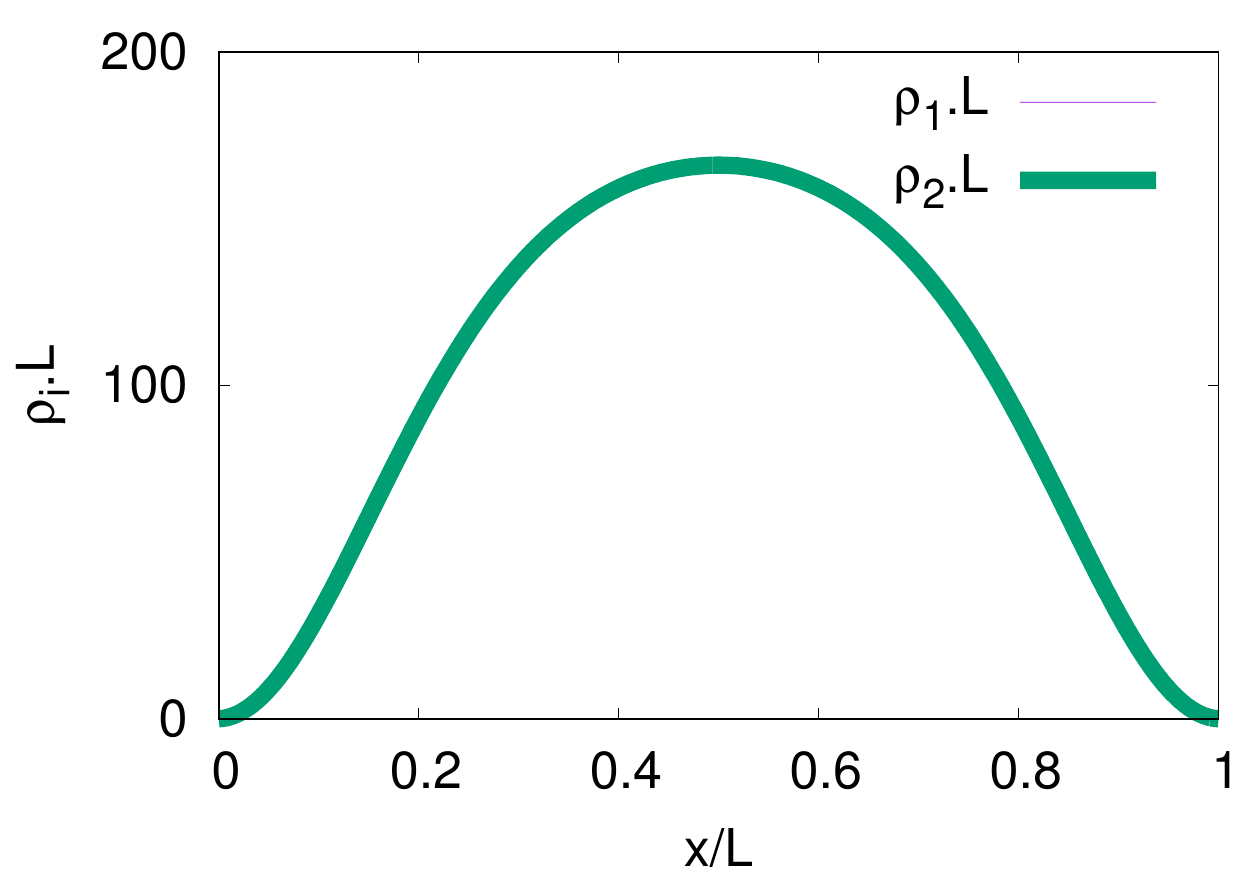}}
\subfigure[]{
\includegraphics[width=0.22\textwidth]{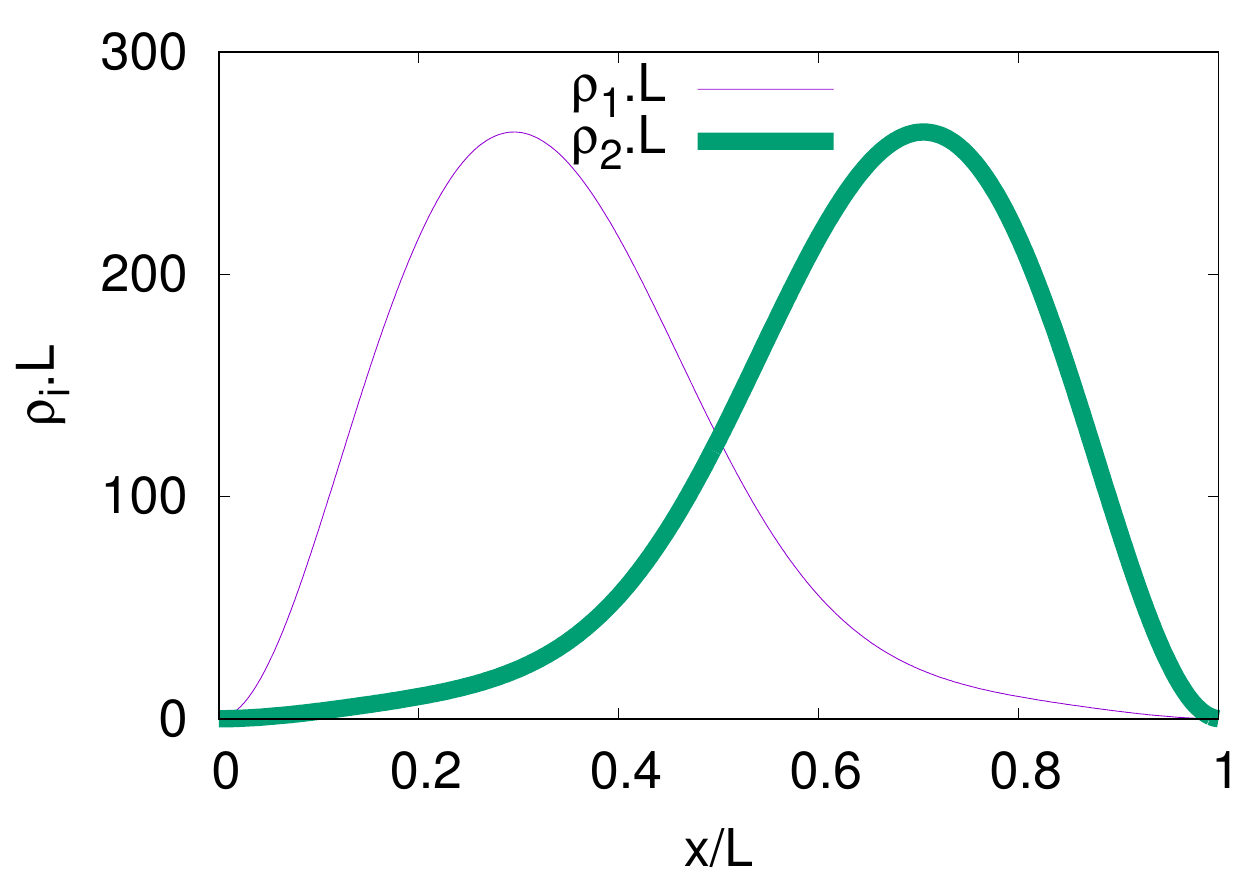}}
\caption{\label{fig:boxsame} Density profiles of equal-mass boson-boson mixtures in a box potential from the imaginary-time GP equations~\eqref{eqn:gp1}. (a) Miscible phase with low inter-species repulsion. $\tilde g_{11}=0.1$, $\tilde g_{22}=0.1$ and $\tilde g_{12}=0.1$. (b) Phase separation due to high inter-species repulsion. $\tilde g_{11}=0.1$, $\tilde g_{22}=0.1$ and $\tilde g_{12}=0.3$. Here $N=100$ for each species and $\tilde g_{ij}=\frac{g_{ij}}{g_0}$.}
\end{figure}

From the two-component GPE for equal-mass boson-boson mixtures, we found a miscible phase when the repulsive inter-species interaction is weak. When the inter-species repulsion in strong, phases separation emerges and the two species congregate into two chunks. Figure~\ref{fig:boxsame} shows the density profiles of typical structures of the equal-mass case. Importantly, the phase separation into two chunks breaks the left-right parity symmetry. We remark that for the equal-mass mixture, the GP equations~\eqref{eqn:gp1} are symmetric between the two components. With a perfect overlap of the initial condition and trap potentials, the system does not enter phase separation if there is no noise during the evolution. The system may enter a fragmented state~\cite{MuellerPRA74_033612} due the the symmetry between the two components. Imperfections in the initial condition or numerical noise in the simulations are enough to drive the system to phase separation in the strongly repulsive regime. Moreover, experiments on equal-mass bosonic mixtures in harmonic traps~\cite{PhysRevLett.81.1539} have shown phase separation.

One reason the few-body and many-body calculations predict different density profiles in the strong repulsion regime is as follows. The many-body wavefunction is a continuous function in space. In the few-body calculations the interactions are spike-like due to the delta-functions, so the wavefunctions of the two species can inter-penetrate without incurring enormous interaction energy. In contrast, the interaction energy of the many-body calculations integrates over the continuous densities and impose a penalty for overlapping wavefunctions. As shown in Fig.~\ref{fig:fewBosonsCorr}, the few-body results suggest ferromagnetic behavior similar to phase separation in the correlations functions. Moreover, imperfections of the initial state and numerical noise have been included in the many-body calculation but not the few-body calculation. If those parity-breaking mechanisms are included in the few-body calculations, the system may also exhibit phase separation in the density profiles. Realistic interactions in experiments may have a finite range, and the dependence of the interaction energy on the wavefunctions may be more complicated.

\begin{figure}[t]\label{fig:ring}
\subfigure[]{
\includegraphics[width=0.22\textwidth]{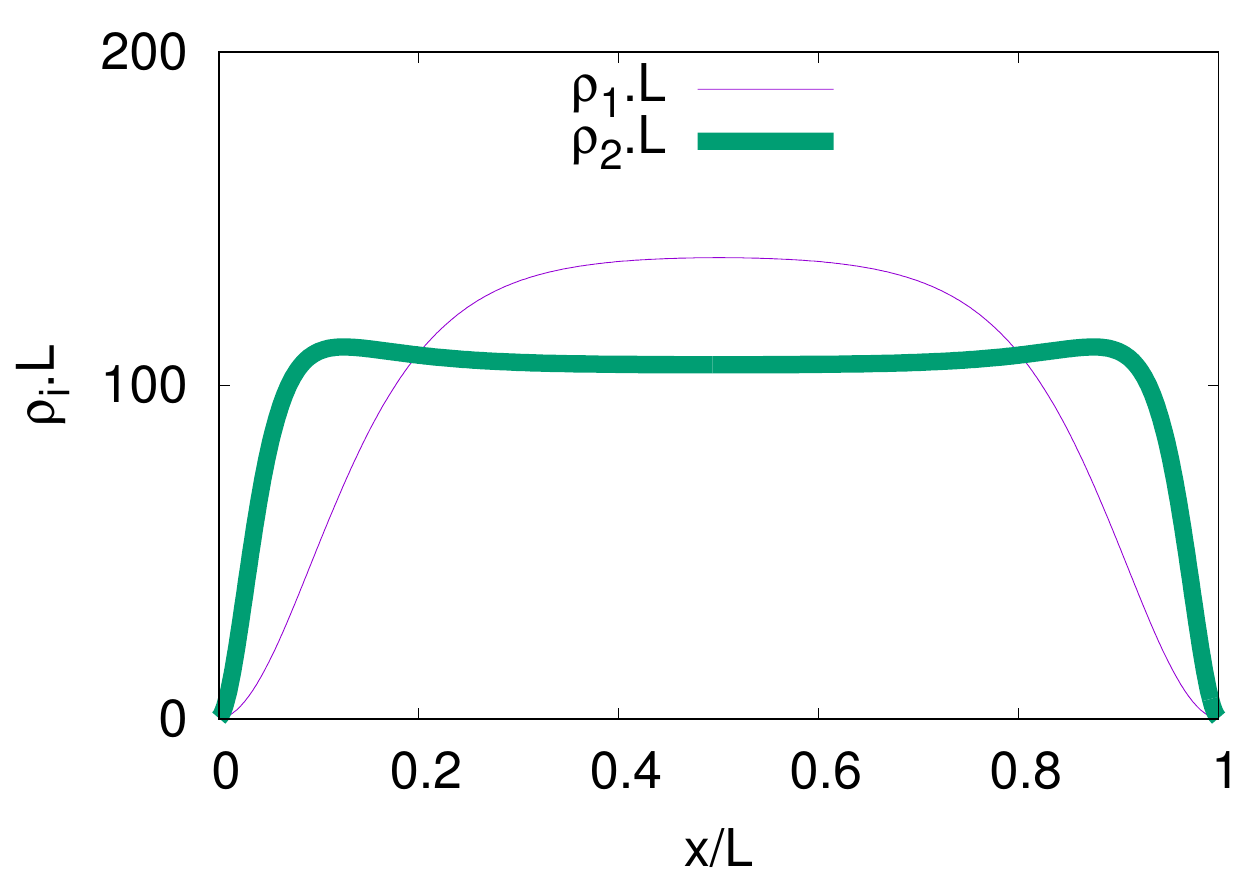}}
\subfigure[]{
\includegraphics[width=0.22\textwidth]{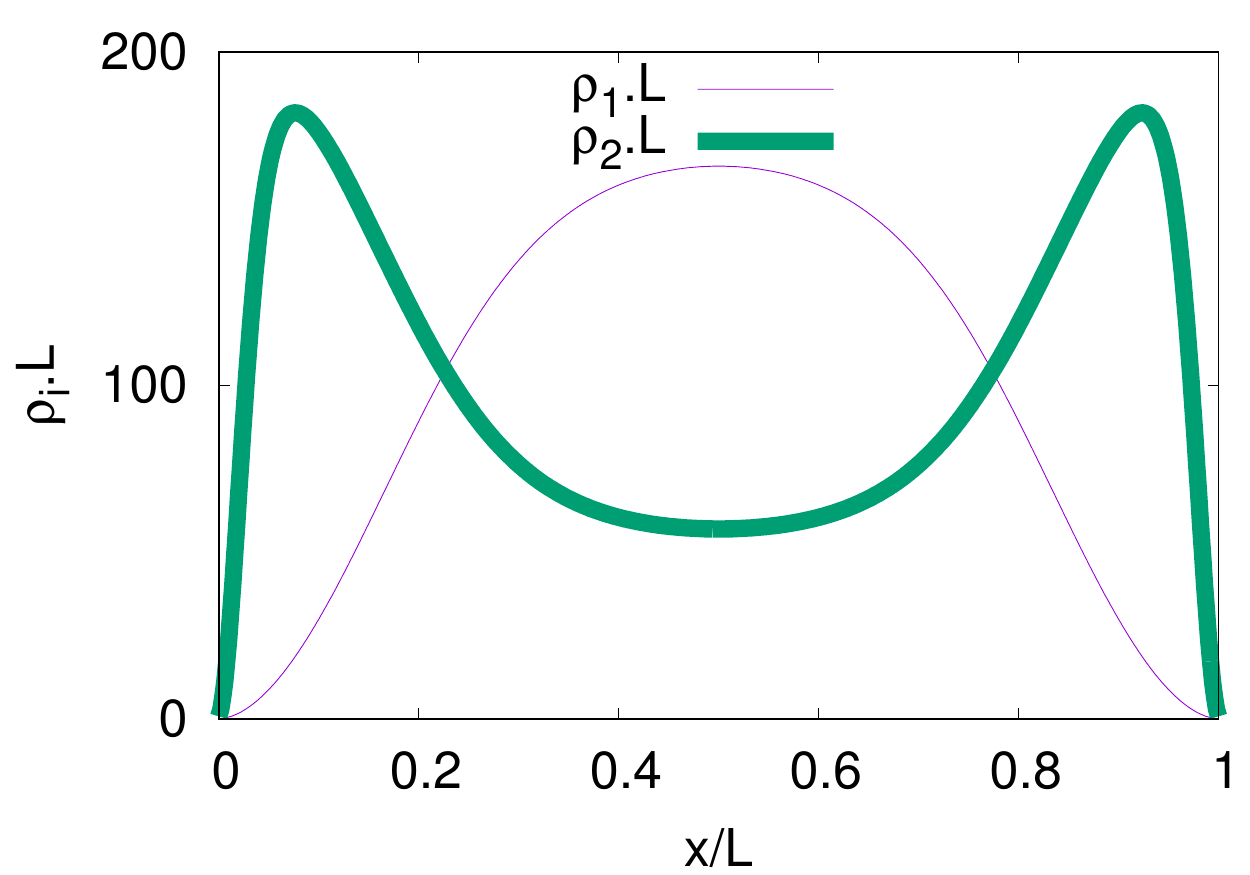}}
\subfigure[]{
\includegraphics[width=0.22\textwidth]{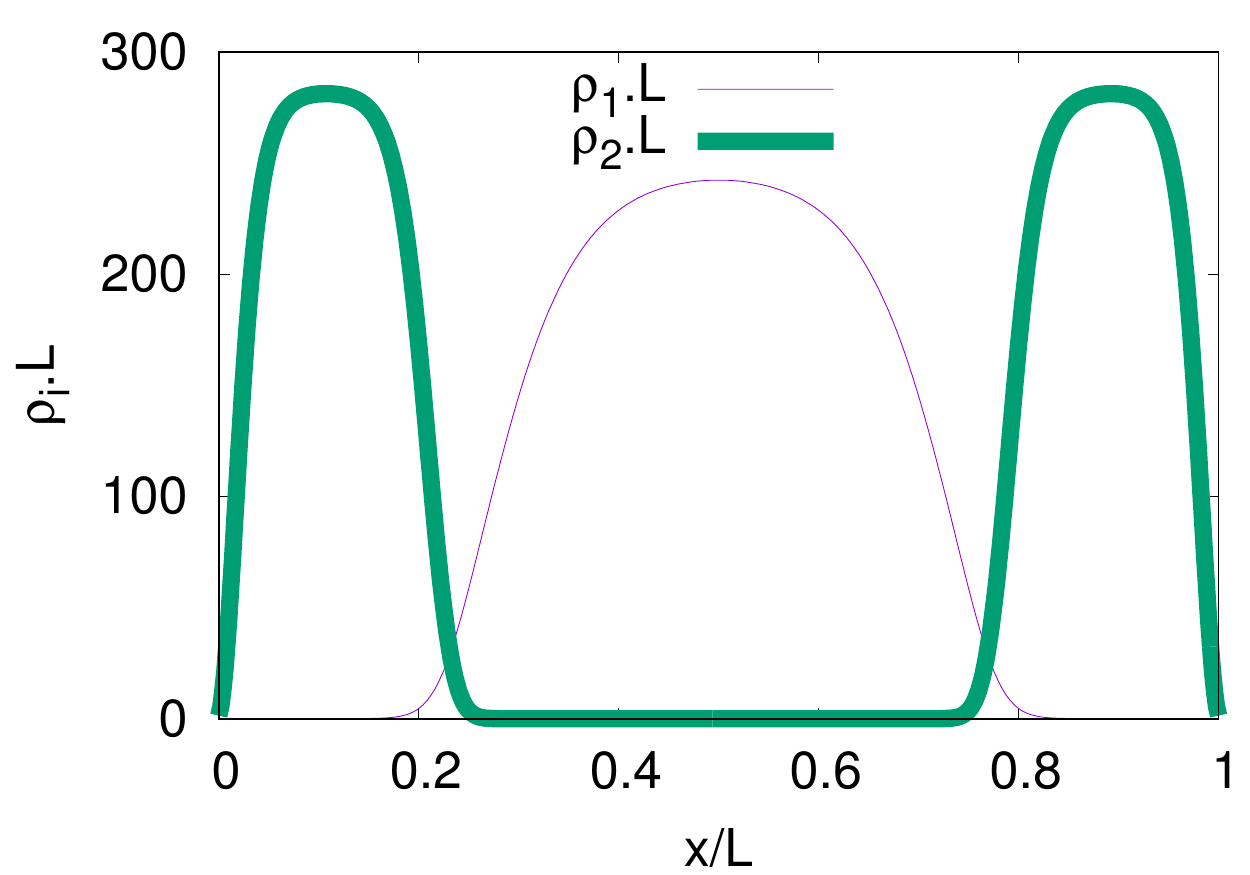}}
\subfigure[]{
\includegraphics[width=0.22\textwidth]{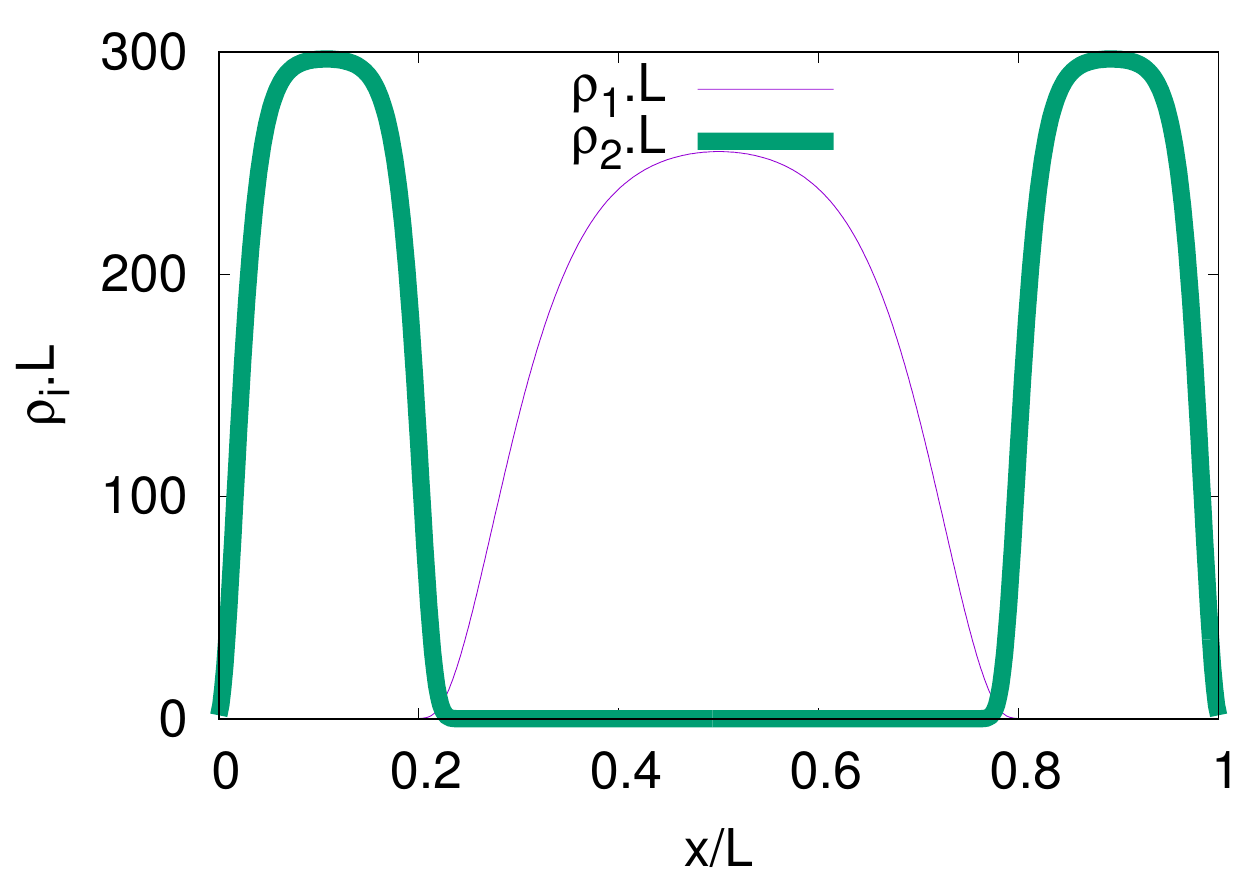}}
\caption{\label{fig:boxdiff} Density profiles of boson-boson mixtures with mass ratio $m_1/m_2=7/87$, showing partial phase separation [(a) and (b)] and full phase separation [(c) and (d)]. Here $N=100$ for each species and $\tilde g_{11}$=$\tilde g_{22}=0.1$ fixed (a) $\tilde g_{12}=0.01$, (b) $\tilde g_{12}=0.1$, (c)  $\tilde g_{12}=1.0$, and (d)  $\tilde g_{12}=20.0$.}
\end{figure}

For boson-boson mixtures with different masses, we take the mixture of $^7$Li and $^{87}$Rb for example. The system exhibits partial and full separation of the two species as the inter-species interaction increases. 
Figure~\ref{fig:boxdiff} (a) and (b) show the partial separation in the intermediate interaction regime while (c) and (d) show the full separation in the strong interaction regime. There is an important difference between the full phase-separation structures of the equal-mass and the different-mass case. For the mass-imbalanced case, the system exhibits a sandwich structure. The lighter species does not touch either of the hard walls. This is because the kinetic-energy increase of the light species at the hard walls can lead to higher total energy, so the lowest-energy configuration has the heavier species locating at both hard-walls. In contrast, there is no such advantage in the equal-mass case, and the two species minimize the number of interfaces between them by separating into only two chunks. As a consequence of the mass imbalance, the phase-separation structure can break the parity symmetry by forming two chunks in the equal-mass case, or the system can keep the parity symmetry in the sandwich structure when the mass imbalance is large.

There are theoretical frameworks for going beyond the GP theory. For example, the auxiliary-field based theory~\cite{PhysRevLett.105.240402} or the method with modified energy functionals~\cite{KolomeiskyPRL85_1146} may be employed to show higher-order correlation effects.

\section{Fermion-fermion mixtures in a box potential}\label{sec:Fmixture}
\subsection{Few-body fermion-fermion mixtures}
We consider a two-component mixture of fermions with a Hamiltonian similar to Eq.~\eqref{eq:ham}. However, the symmetry conditions of the many-body wavefuncions are different in order to satisfy the Bose-Einstein and Fermi-Dirac statistics, respectively.
Next, we decompose the field operator for fermions as  ($\alpha=1,2$)
\begin{equation}
 \hat{\Phi}_\alpha(x) = \sum_i \phi_{\alpha i}(x) \hat{f}_{\alpha i},
\end{equation}
which lead to the following second-quantization Hamiltonian: 
\begin{equation}
 {\cal \hat{H}} = \sum_{\alpha=1}^{2} \sum_i E_{\alpha i}
 \hat{f}_{\alpha i}^\dag \hat{f}_{\alpha i} 
 + g_{12} \sum_{uvkl} U_{uvkl} \hat{f}^\dag_{1l}\hat{f}^\dag_{2k}\hat{f}_{2v}\hat{f}_{1u}.
\end{equation}
The fermions of the same kind do not interact with each other because the Pauli exclusion principle suppresses the two-body $s$-wave interactions between identical fermions.
The fermonic field operators anticommute:
$\{\hat{\Phi}_\alpha^\dag(x),\hat{\Phi}_\beta (x')\}=\delta(x-x')\delta_{\alpha \beta}$.
After the anticommutation relation is properly implemented, the method analogous to the one used for the boson mixture can be utilized. 
To overcome the Fermi pressure, the interaction strength parameter has to be set bigger than in the bosonic case: $g_{12}=50$ to reach strongly correlated regime. Since the computation grows exponentially due to the Hilbert space shown in Table~\ref{tabff} and the convergence of our simulations, we present the results with $N_1=N_2=N=4$ with controllable accuracy and manageable computation time. Moreover, we have checked the parity of the particle number does not change the structures qualitatively.

\begin{figure}
 \includegraphics[width=\columnwidth]{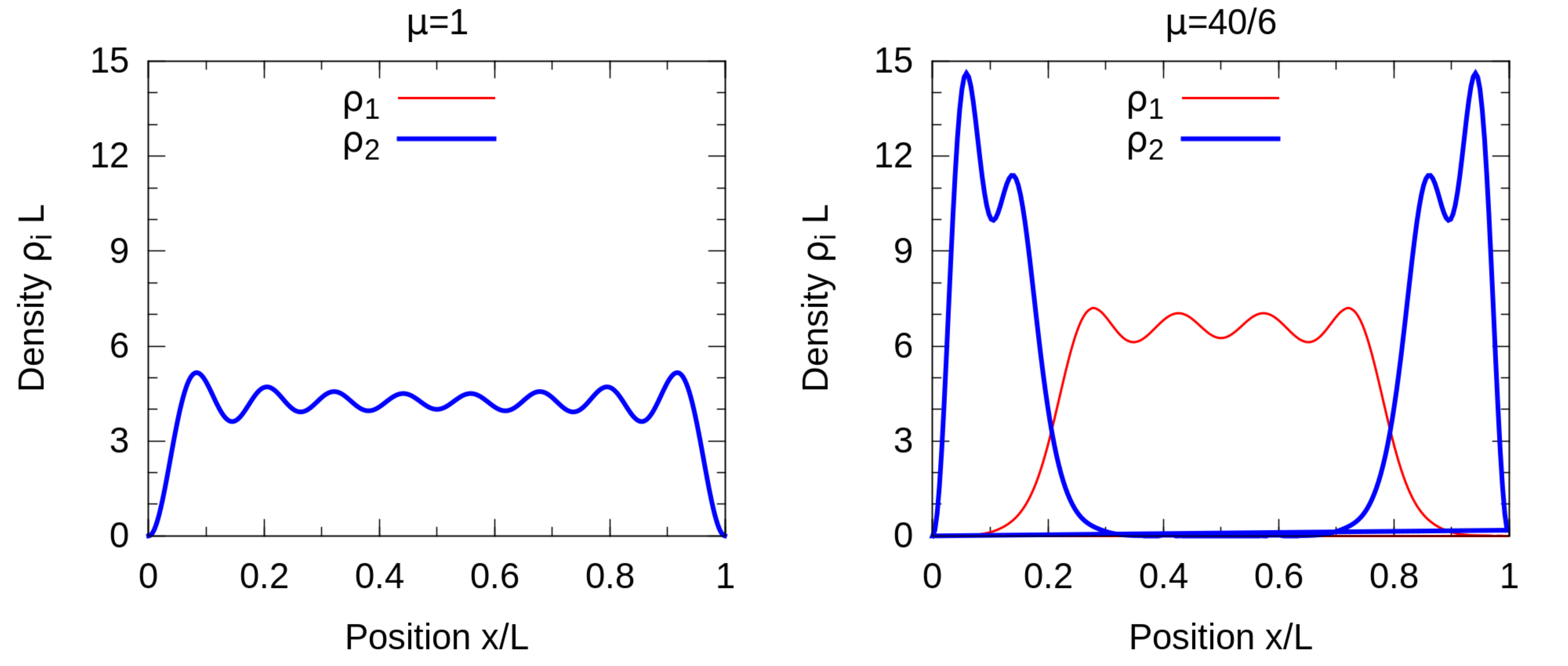}
 \caption{\label{fig:fermionsDensity}Single-particle density profiles for binary fermionic mixtures with $N=4$ particles of each type. The inter-species repulsion is $g_{12}=50$ in the strong repulsion regime. The left (right) panel shows the equal mass case with mass ratio $\mu=1$ ($^{40}$K-$^6$Li mixture with $\mu=m_2/m_1=40/6$).}
\end{figure}

\subsubsection{Equal-mass system}
For an equal-mass fermion mixture, the single-particle density profiles of the two species are exactly the same due to the symmetry in the Hamiltonian. The density distributions in real space minimizing the total energy are shown in the left panel of Fig.~\ref{fig:fermionsDensity} for $N_1=N_2=4$.
There are two mechanisms causing the fermions to avoid each other, leading to the density distributions of the two species.
The first one is the repulsion between different species that lowers the probability of finding two different fermions at the same point in space.
The other is the Pauli exclusion principle forbidding identical fermions from occupying the same quantum state. On the other hand, the delta-function interactions allow the wavefunctions of the two species to inter-penetrate. Therefore, the density profiles show an overlap of the two species, but the two-body correlations shown in the left column of Fig.~\ref{fig:fermionsCorr} exhibit strong anticorrelation  in real space. Specifically, both same-species and different-species correlation functions show that the fermions are avoiding each other.

\subsubsection{Mass imbalanced system}
The presence of the mass imbalance influences the single-particle energy spectrum and makes the distortion of heavy-particle wavefunction more energetically favorable because the single-particle kinetic energy is inversely proportional to the mass. Since the wavefunctions are distorted at the hard walls, the system will lower the total energy by placing the heavy particles there. This implies that the system prefers to make more complicated structures of the heavy particles rather than the light particles.

Therefore, the mass-imbalanced fermion mixture from the few-body calculation exhibits phase separation, showing that the heavy particles are pushed towards the walls and the light particles staying at the center. The miscible structure of the gas is then destroyed in the strong-repulsion regime. Taking a mixture $^6$Li-$^{40}$K as an example, the density profiles from the few-body calculation is shown in the right panel of Fig.~\ref{fig:fermionsDensity}. A three-chunk sandwich structure is observed.

\begin{figure}[t]
 \includegraphics[width=\columnwidth]{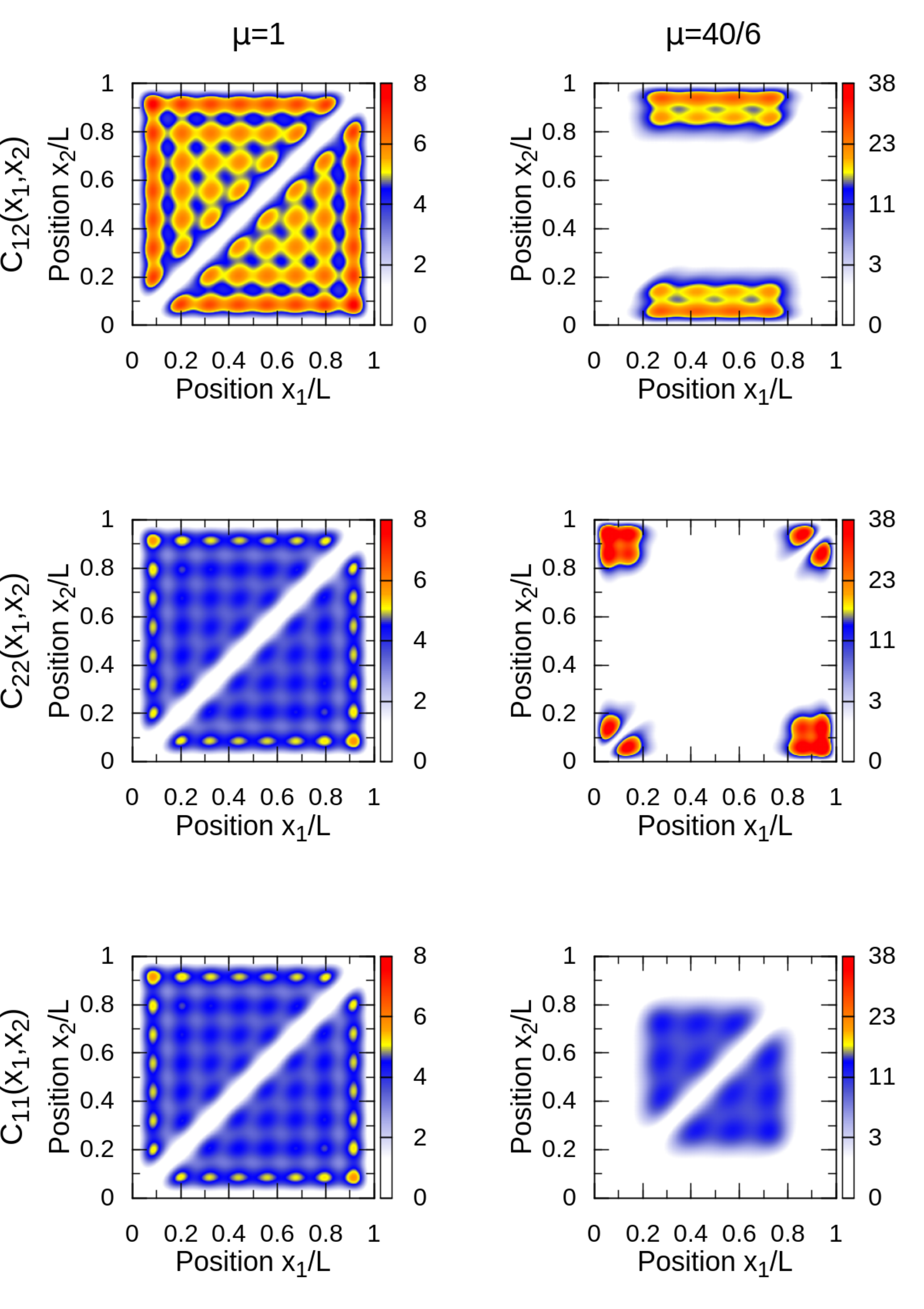}
 \caption{\label{fig:fermionsCorr}Two-particle density correlations $C_{\alpha\beta}$ defined in Eq.~\eqref{eq:Cab} for a system of $N=4$ particles of each type. The inter-species interaction is $g_{12}=50$ in the strong repulsion regime. The  left and right columns show the equal-mass case with mass ratio $\mu=1$ and a $^{40}$K-$^6$Li mixture with $\mu=m_2/m_1=40/6$, respectively.}
\end{figure}

Nevertheless, the mechanisms that lead to the anticorrelation of fermions are still present, implying zero probability of finding a pair of fermions at the same location. The phenomenon can be seen as the vanishing two-body correlations along the diagonals of the plots shown in Fig.~\ref{fig:fermionsCorr}, regardless of mass imbalance. 
The tendency of fermions to avoid each other is seen as the enhanced probability of finding fermions in separate regions rather than at the same spot, which can be observed in the plot of $C_{12}$ shown in  Fig.~\ref{fig:fermionsCorr}. While $C_{12}$ of the equal-mass case shows how the two species inter-penetrate and form staggered correlations, $C_{12}$ for the mass-imbalanced case is only finite at the interfaces of the sandwich structure.

When compared to the equal-mass case, the light-particle correlations are similar but they are confined in the central region due to the sandwich structure.  
In contrast, the heavy-particle correlations are significantly changed due to the spatial separation and the distortion of the wavefunctions at the two hard walls. 
In the presence of mass imbalance, the phase separation structure has the heavy particles confined to two narrow regions near the hard walls.
In systems with $N > 2$, arranging the heavy particles in the two regions near the walls gives rise to finite correlations of finding two heavy fermions around the left or right corner, as seen in the plot of $C_{22}$ in Fig.~\ref{fig:fermionsCorr}.

\subsection{Many-body theory and result}
Similar to the many-body approach to the bosonic mixtures, we model the interactions between fermions as density-density contact interactions except there is virtually no intra-species interaction between identical ultracold fermionic atoms due to the Pauli exclusion principle. The second-quantization Hamiltonian of a fermion-fermion mixture with repulsive interactions is given by
\begin{eqnarray}\label{eq:HF}
\hat{H}_F&=&\int d^{d}x \Big(\sum_{\alpha=1}^2\frac{\hbar^2}{2m_\alpha}|\nabla \hat{\psi}_\alpha|^2+g_{12}\hat{\rho_1}\hat{\rho_2}\Big).
\end{eqnarray}
Here $m_\alpha$ is the mass of the $\alpha$-th species, $\hat{\rho}_{\alpha}=\hat{\psi}^{\dagger}_{\alpha}\hat{\psi}_{\alpha}$ is the density operator, $\alpha=1,2$.

\subsubsection{Equations for fermion-fermion mixtures}
By using the Hartree-Fock approximation~\cite{Fetter_book}, we replace the interaction term by the expectation values $\rho_\alpha=\langle \hat{\rho}_\alpha \rangle$. Assuming the hard-wall confinement is along the $x$-direction, we obtain the following eigenvalue equations for the two species consistent with the stationary states of Eq.~\eqref{eq:HF}.  
\begin{eqnarray}\label{eqn:fermi}
    -\frac{\hbar^2}{2m_1}\frac{\partial^2}{\partial x^2}\psi_{1,n} +  g_{12}\rho_2\psi_{1,n}&=& E_{1,n}\psi_{1,n}, \nonumber \\
    -\frac{\hbar^2}{2m_2}\frac{\partial^2}{\partial x^2}\psi_{2,n} +  g_{12}\rho_1\psi_{2,n}&=& E_{2,n}\psi_{2,n},
\end{eqnarray}
where $E_{\alpha,n}$ are the eigenvalues corresponding to the eigenstates $\psi_{\alpha,n}$, which are from the decomposition of $\hat{\psi}_\alpha$.
The boundary conditions are $\psi_{i,n}=0$ at $x=0,L$.
We choose $m_1$, $L$, and $\hbar^2/(2m_1 L^2)$ as the units of mass, length, and energy, respectively. In the ground state, fermions of species $\alpha=1,2$, occupy the lowest $N_\alpha$ levels of $\psi_{\alpha,n}$. As a consequence,  $\rho_\alpha=\sum_{n=1}^{N_\alpha}|\psi_{\alpha,n}|^2$. Thus, the densities satisfy
\begin{eqnarray}
N_{\alpha}=\int_0^L dx \rho_\alpha,
\end{eqnarray}
$\alpha=1,2$. We will use $N_1=N_2=100$ in our illustrations.

For a given value of $g_{12}$, we use an iteration method similar to the one for solving the Bogoliubov-de Gennes equation of superconductivity~\cite{zhu2016bogoliubov}.  We start with a set of trial density profiles and solve the coupled equations~\eqref{eqn:fermi} to obtain the eignevalues and their normalized eigenvectors. From the eigenvectors, we calculate the iterated densities $\rho_1$ and $\rho_2$ and use them to solve the coupled equations~\eqref{eqn:fermi} again. The iteration converges if  ($\int_{0}^{L}|\rho_1^{\nu+1}-\rho_1^\nu|dx<\epsilon$ and $\int_{0}^{L}|\rho_2^{\nu+1}-\rho_2^\nu|dx<\epsilon$) for the $\nu$-th iteration.
In our calculation, we choose $\epsilon=10^{-3}$ and $10^4$ grid points in $x/L\in[0,1]$. We have verified that further changing the tolerance or grid size does not lead to qualitative change as long as the grid number is much larger than the particle number, the latter condition sets a limitation of our many-body calculations. We remark that the mean-field, many-body treatment of fermions overlooks the correlation effect, and the zero-range contact interaction is not accurately incorporated. Moreover, Eqs.~\eqref{eq:HF} and ~\eqref{eqn:fermi} assume continuous wavefunctions and smooths out possible local structures.

\begin{figure}
\subfigure[]{
\includegraphics[width=0.22\textwidth]{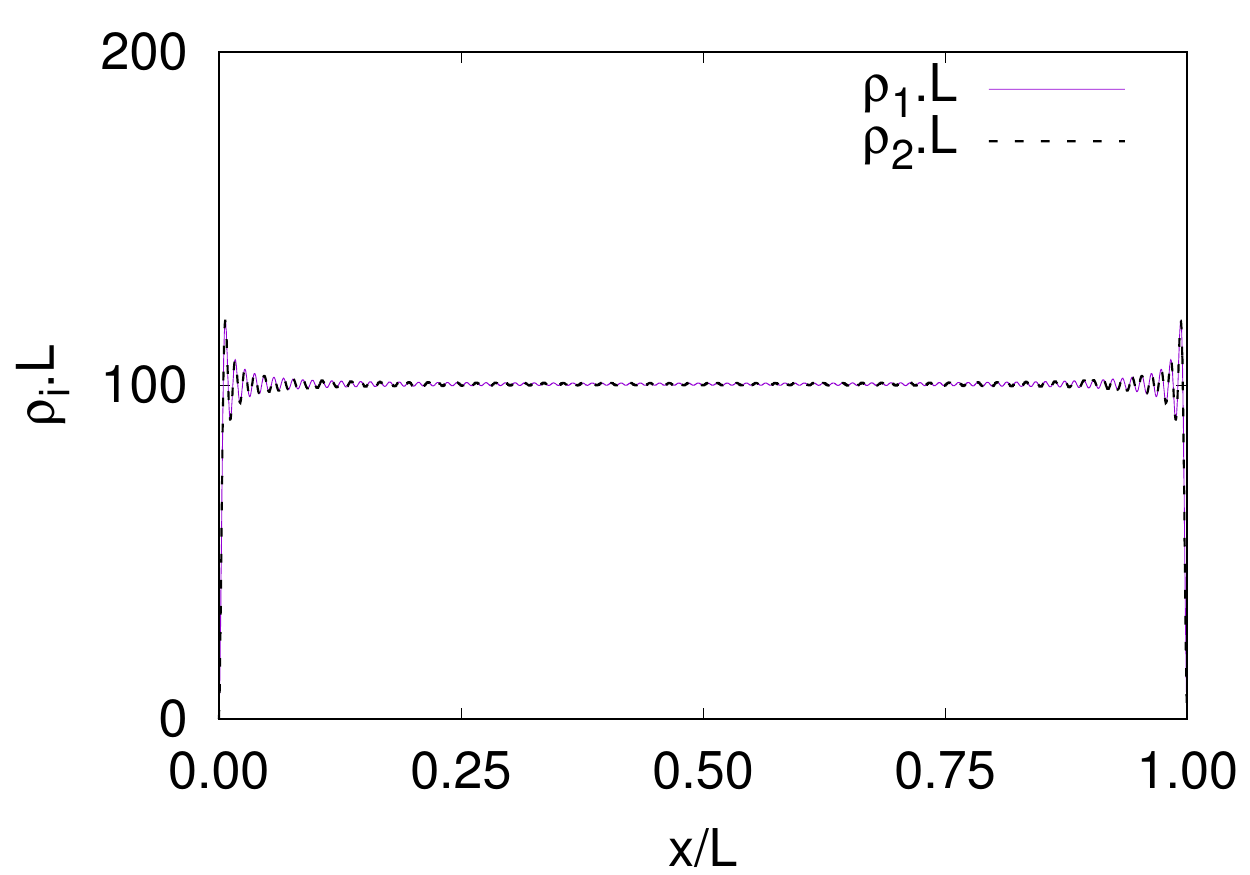}}
\subfigure[]{
\includegraphics[width=0.22\textwidth]{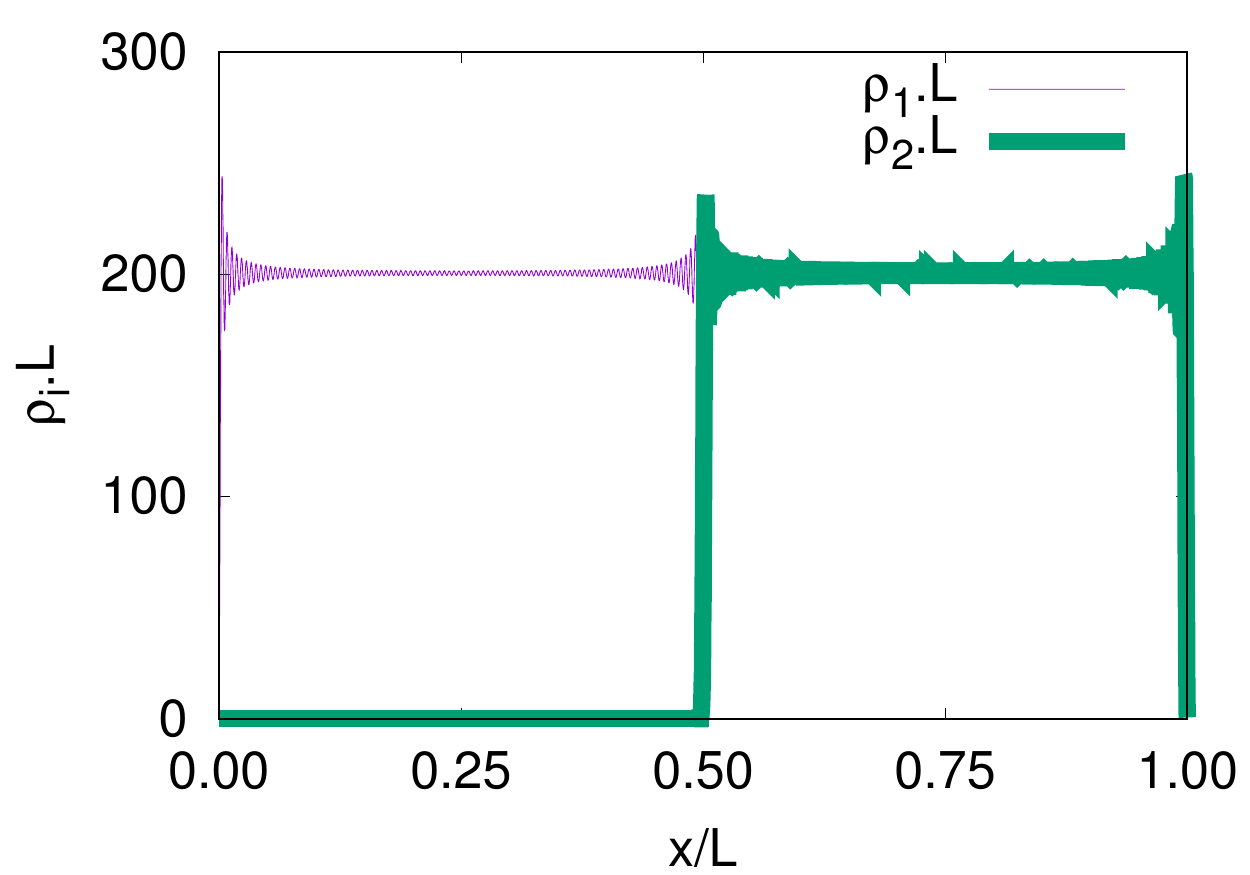}}
\subfigure[]{
\includegraphics[width=0.22\textwidth]{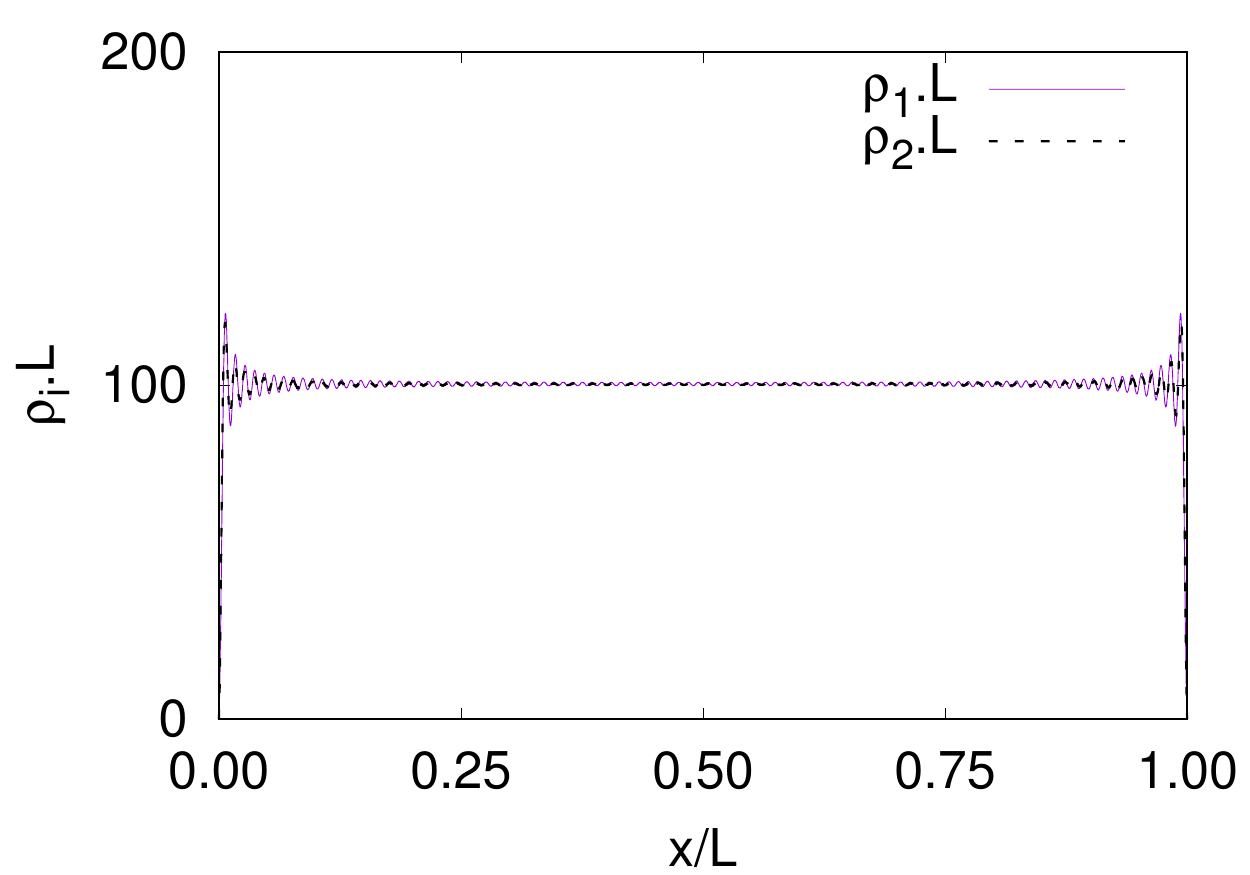}}
\subfigure[ ]{
\includegraphics[width=0.22\textwidth]{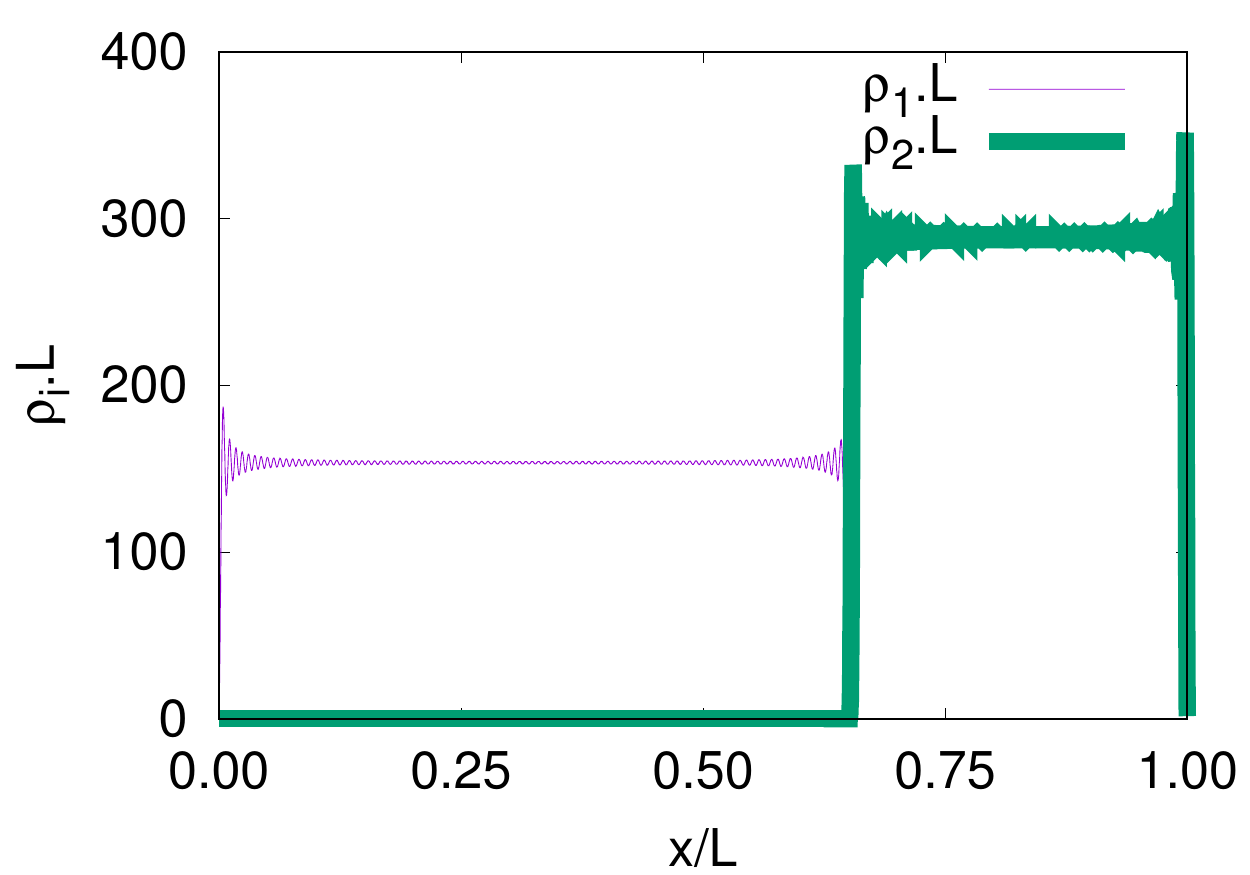}}
\subfigure[ ]{
\includegraphics[width=0.22\textwidth]{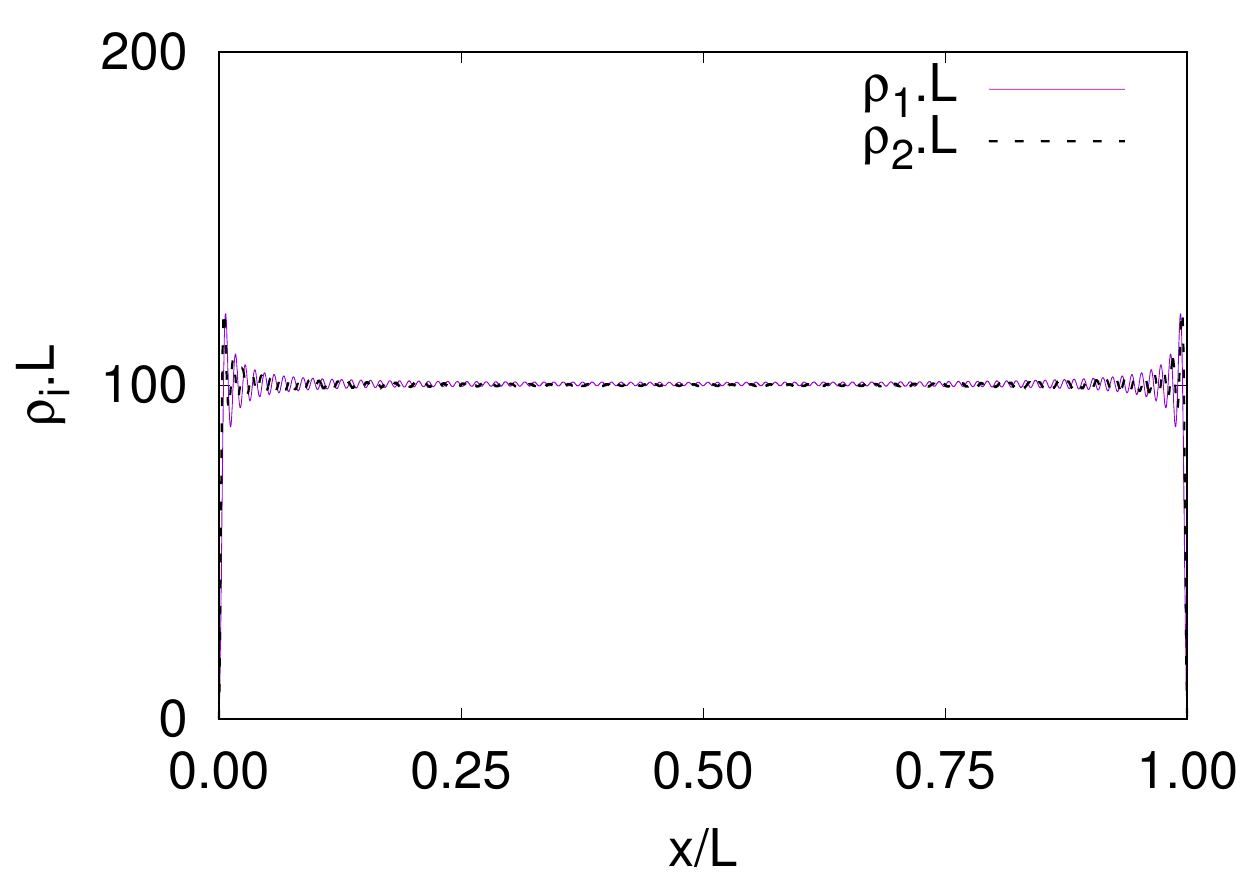}}
\subfigure[ ]{
\includegraphics[width=0.22\textwidth]{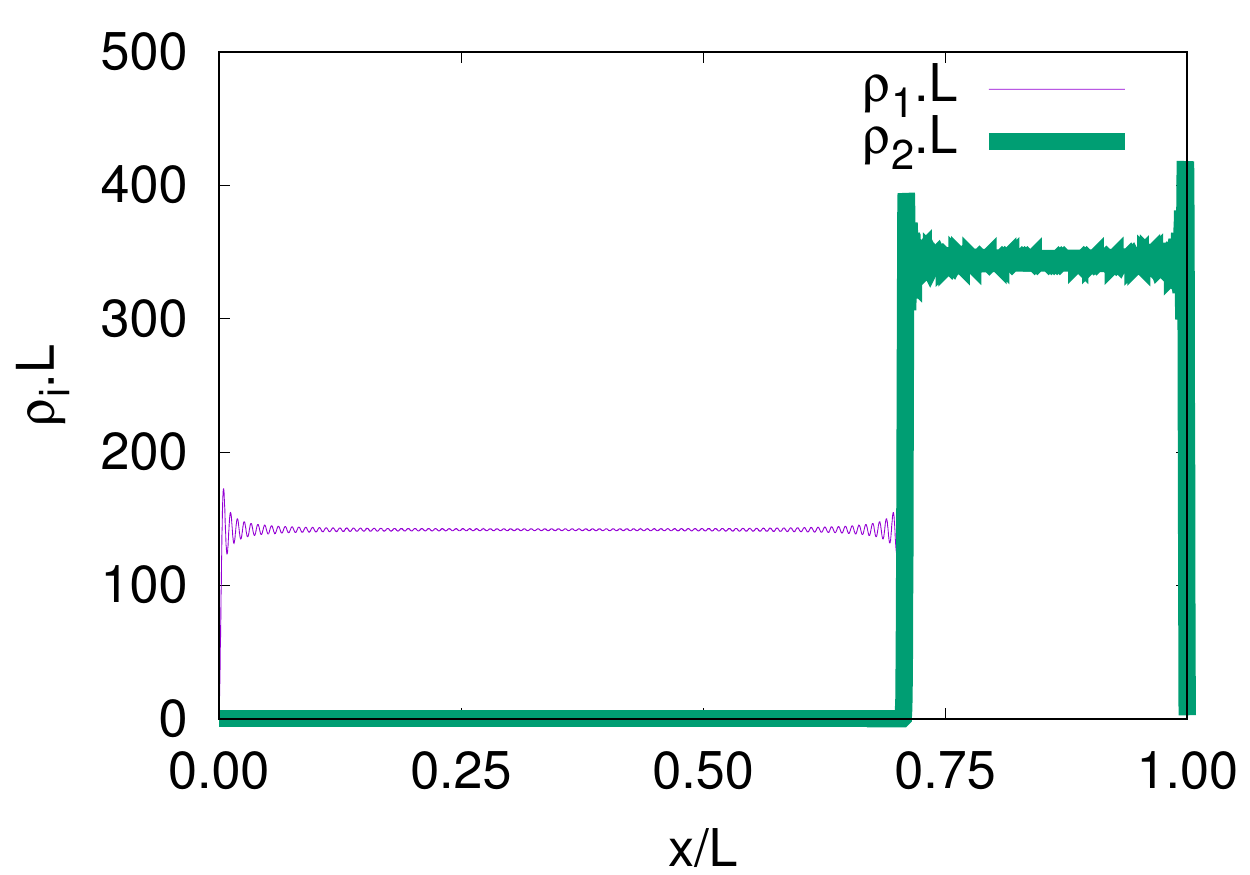}}
    \caption{\label{fermion}Density profiles of fermion-fermion mixtures in 1D box potentials for the case with equal mass shown in (a) and (b), for the case with mass ratio $m_1/m_2=6/40$ shown in (c) and (d) and for the case with mass ratio $m_1/m_2=6/86$ shown in (e) and (f). The left column [(a),(c), and (e)] shows the miscible structures in the weak repulsion regime, and the right column [(b),(d), and (f)] shows the phase separation structures in the strong repulsion regime. Here $N_1=N_2=100$ and the values of $\tilde g_{12}$ are (a) $300$, (b) $5000$, (c) $10$, (d) $500$, (e) $10$, and (f) $500$.}
\end{figure}

\subsubsection{Results} 
Figure \ref{fermion} shows the structures of fermion-fermion mixtures from the Hartree-Fock theory. In the many-body calculation, the Fermi pressure from the kinetic energy is a dominant factor. The system tends to remain in the miscible phase to minimize the kinetic energy until the repulsion is strong enough to overpower the Fermi pressure.

For the equal-mass case shown in Figure \ref{fermion} (a) and (b), the system phase separates into two chunks when the inter-species repulsion overpowers the bulk kinetic energy. The phase separation is a consequence of the density-density interactions because the two species tend to occupy different spatial regions to minimize the overlap of their densities. For the delta-function interaction used in few-body systems, the wavefunctions can inter-penetrate each other without incurring severe interaction-energy penalty, and the left panel of Fig.~\ref{fig:fermionsDensity} shows no spatial phase separation in the few-body case even in the strong repulsion regime. Since the density-density interaction and the delta-function interaction are idealized models of the real interactions between atoms, experimental results may depend on the details of the two-body scattering as well as the density of the cloud. We remark that for bosonic or fermionic mixtures, by reducing the particle numbers of the many-body calculations to be below $10$, the results start to show features of the few-body results such as density modulations. However, the density profiles are still different, showing the breakdown of mean-field approximation in the few-body limit.

To test the mass-imbalance effect, we consider fermion mixtures with larger mass ratios.
The phase separation structures of $^6$Li - $^{40}$K and $^6$Li - $^{86}$Rb mixtures from the many-body calculations are shown in Fig.~\ref{fermion} (d) and (f), respectively. Due to the strong repulsive interactions countering the high kinetic energy from the separation, the kinetic energies at the boundary due to the distortion of the wavefunctions do not play an important role. Thus, the system forms two chunks in phase separation similar to the equal-mass case. This is in contrast to the few-body case with delta-function interactions, where a sandwich structure is shown in the right panel of Fig.~\ref{fig:fermionsDensity}. For the few-body case, an intermediate mass ratio ($m_2/m_1=40/6$) already leads to the sandwich structure. In contrast, the bulk kinetic energy of a many-body system is large in the phase separation regime, so the mass imbalance induced kinetic energy change at the boundary does not play a significant role. Even with a large mass ratio of $m_2/m_1=86/6$, no sandwich structure is observed in the many-body calculation. We mention that (1) a further increase of the interactions with a fixed mass ratio adds more bulk energy to the system than the interfacial energies at the boundaries, so there is still no sandwich structure; (2) it is possible to introduce extremely large mass imbalance in the many-body calculations to produce a sandwich structure, but the parameters are inaccessible to cold-atom experiments due to the limited atomic species.

The structures of fermion-fermion mixtures with tunable repulsive interactions in a 1D box potential thus depend on many factors, including how the atomic interactions are modeled, the mass ratios, and how many particles are confined in the box. Those factors are in principle tunable in experiments, and the various structures may be realized with suitable conditions. We mention in the ultra-strong interaction regime, one-dimensional fermions may be approximated by spin-chains~\cite{DeuretzbacherPRA90_013611,VolosnievPRA91_023620} when the kinetic energy is a perturbation to the interaction energy. Our results, however, is in the regime where the kinetic and interaction energies are comparable and the systems enter phase separation structures. We also comment that the many-body results of bosons and fermions in the strong-interaction regime already exhibit phase separation structures, unlike the few-body results where the density profiles may not reveal how the two species repel each other. Therefore, the two-particle correlation plots offer complementary information to the few-body calculations but not the many-body calculations.

\section{Experimental implications}\label{sec:exp}
The mass ratio of available atomic mixtures is not a continuously tunable parameter, and it is limited by the available species in the periodic table. Moreover, the inter-species and intra-species interactions are not independently controllable in cold-atom experiments because tuning the magnetic field will adjust the values of the inter- and intra- species two-body scattering lengths~\cite{pethick2008bose}. 
However, the structures from the few-body and many-body calculations with reasonable values of the mass ratio and interaction strengths show that the structures of atomic mixtures can be different with or without mass imbalance due to the box potential. We remark that while the structural difference in the box potential due to mass-imbalance is caused by local kinetic energy change at the hard walls, the structural difference in harmonic traps are due to a global competition of kinetic energy and potential energy. For many-body equal-mass bosonic mixtures, noise and imperfection in experiments help break symmetry between the two components and favor phase separation, similar to the case of harmonically trapped bosonic mixtures~\cite{PhysRevLett.81.1539}.

The box potentials in Refs.~\cite{Es10,Gaunt2013,PhysRevLett.112.040403,PhysRevLett.118.123401,Horikoshi17,PhysRevLett.120.060402} are 3D traps. If the argument based on the competition between kinetic and interaction energies works for higher-dimensional systems, the distortion of wavefunctions at the hard walls leads to additional kinetic energy and may affect the structures of atomic mixtures in the presence of mass imbalance. For an equal-mass bosonic mixture, a miscible phase with strong internal correlations for the few-body case and a two-chunk phase separation for the many-body case in the strong repulsion regime are expected. However, in the presence of mass imbalance, the light species tends to avoid the hard walls to lower the kinetic energy due to wavefunction distortion. In a 3D box, a strong mass imbalance may lead to a core of the light particles enclosed by the heavy particles to lower the kinetic energy at the hard walls. Such a possibility has been overlooked in analyses performed in the thermodynamic limit where the boundary does not play a role~\cite{pethick2008bose,PhysRevLett.81.5718,JasonHo2013PhaseSeparation}.

While the few-body result of binary fermionic mixtures with mass imbalance shows possible enclosure of the light particles by the heavy ones, the dominating bulk kinetic energies in the many-body case favors two-chunk phase separation for reasonable range of mass imbalance. However, the structures of fermionic mixtures are further complicated by other instabilities~\cite{Jo09,Pekker11,DeuretzbacherPRA90_013611,VolosnievPRA91_023620}, so the system may enter other phases or cannot be trapped in the strongly interacting regime. 

Atomic clouds have been cooled down to the extent where thermal excitations do not play a significant role~\cite{pethick2008bose,Es10,Gaunt2013,PhysRevLett.112.040403,PhysRevLett.118.123401,Horikoshi17}. If thermal behavior is of interest, it is possible to incorporate finite-temperature effects in many-body calculations using the Bogoliubov-de Gennes method~\cite{pethick2008bose} or large-N expansion~\cite{ChienPRA12} for bosons or including the Fermi distribution function in the self-consistent equations for fermions~\cite{zhu2016bogoliubov}. The kinetic energy increases with temperature and is expected to suppress the regime where phase separation can be observed.
However, Ref.~\cite{NJP013030} shows phase separation at finite temperatures using few-body calculations of fermions in a harmonic trap. As the total kinetic energy increases with the particle number at finite temperatures, the threshold for entering phase separation in many-body systems will increase accordingly.

\section{Conclusion}\label{sec:Conclusion}
From the few-body calculations with delta-function interactions and many-body calculations with density-density interactions, we have shown that mass imbalance can lead to different structures of boson-boson mixtures in 1D box potentials when compared to the equal-mass case. The structural difference comes from the competition between the interaction energy and kinetic energies due to the density distortion at the hard walls and the phase-separation interface. For fermion-fermion mixtures, different structures with and without mass imbalance were found in the few-body calculations in the strong repulsion regime. For many-body fermion mixtures to enter the phase-separation regime, however, the bulk kinetic energies increase due to the reduced volume of each species in phase separation. Therefore, many-body calculations of fermion mixtures in the phase separation regime show a two-chunk structure within the reasonable range of mass imbalance. The few-body and many-body calculations sometimes predict different structures for bosonic or fermionic mixtures because of the different models of the interactions. Depending on the experimental conditions, one may check if the structures of atomic mixtures resemble the many-body or few-body results.

Some related topics may be explored in the future:
While binary boson-fermion mixtures are also realizable in cold atoms~\cite{Onofrio04}, their theoretical treatments are complicated~\cite{Minguzzi04,Tom18,karpiuk2004soliton}. Nevertheless, their structures in box potentials may be analyzed in a similar framework. Moreover, dynamics of atomic mixtures across phase separation transitions has been studied~\cite{ChienQuench13,MistakidisNJP18,Kiehn19}. The mass-imbalance effect may be incorporated to enrich the physics.

\acknowledgements
This work was supported by the Polish National Science Center (NCN) under Contract No. UMO-2017/27/B/ST2/02792 (DP).
Numerical calculations were partially carried out in the Interdisciplinary Centre for Mathematical and Computational Modeling, University of Warsaw (ICM) under the computational grant No. G75-6.

\appendix 
\section{Numerical convergence of few-body calculations}\label{appendix}
\begin{figure}
 \includegraphics[width=0.49\textwidth]{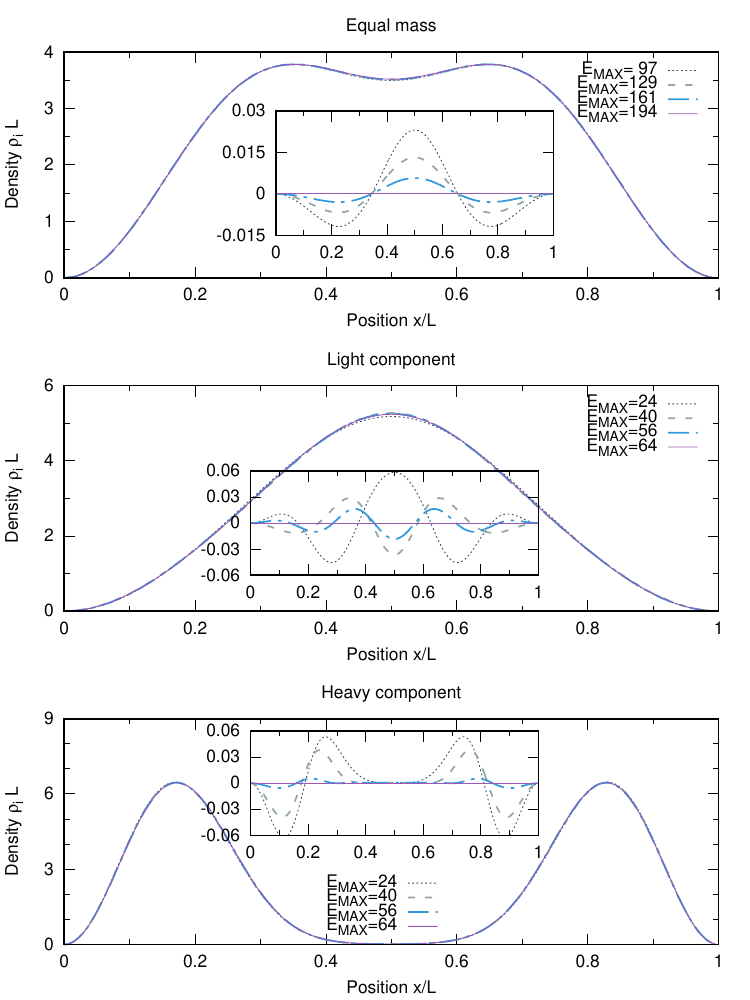}
\caption{The single-particle density of a boson-boson mixture with different many-body cutoffs: The equal-mass case (top), light component (middle), and heavy component (bottom) of a $^{87}$Rb-$^{7}$Li mixture. The insets show the density difference from the density profile with the maximal cutoff of each panel ($E_{MAX}=194$, $64$ and $64$ for the top, middle and bottom panels, respectively).
\label{fig:BBconv}}
 \end{figure}
 
\begin{figure}
 \includegraphics[width=0.49\textwidth]{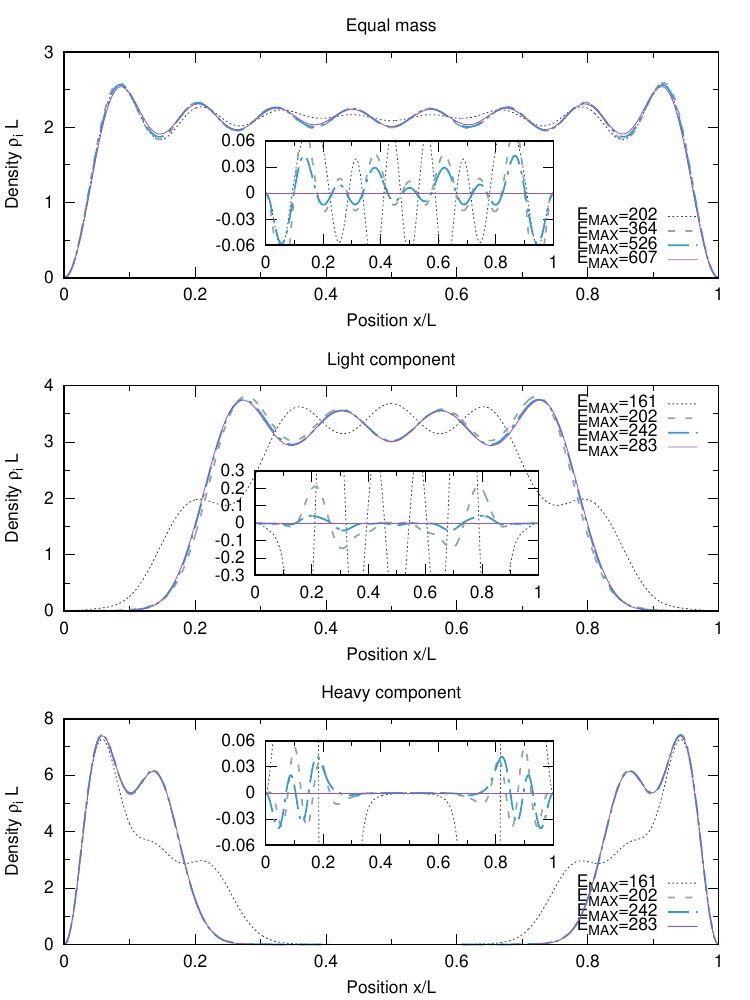}
 \caption{The single-particle density of a fermion-fermion mixture with different many-body cutoffs: The equal-mass case (top), light component (middle), and heavy component (bottom) of a $^{40}$K-$^{6}$Li mixture. The insets show the density difference from the density profile with the maximal cutoff of each panel ($E_{MAX}=607$, $283$ and $283$ for the top, middle and bottom panels, respectively).
 \label{fig:FFconv}}
\end{figure}

We check the numerical convergence of the few-body calculations by systematically varying the cutoff. 
If the observables do not change significantly, we assume that the convergence has been achieved. 
In this paper we focus on the phase separation in the single-particle densities, thus this is the observable that is checked for the convergence.
Here, we show how the results change with the many-body energy cutoff for boson-boson and fermion-fermion mixtures in Fig.~\ref{fig:BBconv} and Fig.~\ref{fig:FFconv}, respectively. The chosen values are already high enough for the densities to converge. The energy is measured in unit of the single-particle ground-state energy of the lighter species, $\hbar^2\pi^2/(2 m_1 L^2)$.

It is worth noting that the cutoff for mass-imbalanced systems is much smaller, due to the fact that the single-particle energy of a heavy particle is smaller than the light one (see Eq.~\eqref{eq:spHam}). 
Nevertheless, the dimensions of the Hilbert spaces of equal-mass and mass-imbalanced mixtures remain comparable.
When the mass-imbalance is large, there are essentially two energy scales in the system. This explains that the excitations among the heavy species are favored, leading to a push of the heavy particles out to the boundaries.
From a technical point of view, the lighter species can be accurately described with much fewer single-particle states than the heavier species.
Finally, the cutoff for the fermion-fermion mixture is higher compared to the boson-boson case due to the Pauli exclusion principle.

\bibliographystyle{apsrev4-1}
%

\end{document}